\documentclass[prd,twocolumn,nofootinbib]{revtex4}

\usepackage{graphicx,latexsym,amsfonts,amsmath,amssymb,mathbbol,slashed}

\graphicspath{{./figures/}}

\def\diag{{\rm diag}}
\def\tr{{\rm tr}\,}
\def\Pol{{\mathcal{P}}}
\def\Hol{{\mathcal{P}_\infty}}
\def\IPR{\textnormal{IPR}}
\def\half{\frac{1}{2}}
\def\unity{\mathbb{1}}
\def\Re{{\rm Re}}
\def\Im{{\rm Im}}
\def\nn{\nonumber}
\def\noi{\noindent}

\newlength{\fs}
\newcommand{\beq}{\begin{equation}}
\newcommand{\eeq}{\end{equation}}
\newcommand{\beqs}{\begin{equation*}}
\newcommand{\eeqs}{\end{equation*}}
\newcommand{\bea}{\begin{eqnarray}}
\newcommand{\eea}{\end{eqnarray}}
\newcommand{\beas}{\begin{eqnarray*}}
\newcommand{\eeas}{\end{eqnarray*}}
\newcommand{\mref}[1]{Eq.~(\ref{#1})}
\newcommand{\fref}[1]{Fig.~\ref{#1}}
\newcommand{\tref}[1]{Tab.~\ref{#1}}
\newcommand{\sref}[1]{Section~\ref{#1}}
\newcommand{\tx}[1]{\textnormal{#1}}
\newcommand{\mc}[1]{\mathcal{#1}}

\begin{document}
\thispagestyle{empty}

\date{\today}
\title{Calorons in $SU(3)$ lattice gauge theory}
\author{E.-M. Ilgenfritz}
\author{M. M\"uller-Preussker}
\author{D. Peschka}
\affiliation{Humboldt-Universit\"at zu Berlin\\ 
Institut f\"ur Physik, Newtonstr.~15\\
12489 Berlin, Germany}

\begin{abstract}
We examine the semiclassical content of $SU(3)$ Yang Mills theory on
the lattice at finite temperature. Employing the cooling method, a set
of classical fields is generated from a Monte Carlo ensemble. Various
operators are used to inspect this set with respect to topological
properties. We find pseudoparticle fields, so-called caloron
solutions, possessing the remarkable features of (superpositions of)
Kraan-van Baal solutions, i.e. extensions of Harrington-Shepard
calorons to generic values of the holonomy.
\end{abstract}

\maketitle

\section{Introduction}\label{intro}
Guided by the idea that the semiclassical approach to 
QCD \cite{'tHooft:1976fv,Callan:1977gz,Coleman:1978ae}
provides a reasonable model for the QCD vacuum, one first has to 
specify the basic set of classical solutions of the underlying  
Yang-Mills theory. Mathematical rich and interesting structures
naturally arise in the construction and topological classification of
these fields. 
For $SU(N)$ gauge theory in the Euclidean space $\mathbb{R}^4$
solutions with finite action are the well known BPST instantons
\cite{Belavin:1975fg} and their generalizations to arbitrary
topological charge \cite{Atiyah:1978ri}.  
The vigorous work pursuing the semiclassical approach culminated in
the so-called instanton-liquid model \cite{Ilgenfritz:1980vj,%
Shuryak:1981ff,Diakonov:1983hh}, which successfully realizes
fundamental non-perturbative properties of QCD such as the $U(1)_A$
anomaly and chiral symmetry breaking. It has led to numerous
phenomenological  applications (for newer reviews see
\cite{Schafer:1998wv,Diakonov:2002fq}). However, it failed to
reproduce one essential feature of QCD -- confinement.%
\footnote{This incited the work of Negele et al., who
  were able to reproduce confinement in a gas of regular
  gauge instantons or merons \cite{Negele:2004hs}.}

In case of the $SU(N)$ gauge theory with finite temperature defined on 
$S^1 \times \mathbb{R}^3$ quark (de)confinement is closely related 
to the global center symmetry. The Polyakov loop at temperature $~T~$ 
\beq
\Pol(\vec{x})=
{\rm P}\exp\left(\int_0^{b=1/T}\!\!\!A_0(x_0,\vec{x})\,dx_0\right) \; ,
\eeq
an order parameter of this symmetry, signals a phase
transition from a deconfining to a confining phase. Deconfinement is a
feature of the high temperature phase, where the symmetry is spontaneously 
broken and -- in the limit of large temperatures -- the Polyakov loop takes  
values in the center of the gauge group $~Z(N)$. Below the critical 
temperature the symmetry becomes restored, the trace of the Polyakov loop 
fluctuates around zero and confinement is observed. 
One can hope that not only for temperatures above the transition temperature
the semiclassical approach is justified. Then it should be able to reflect 
the transition as an important feature of the finite temperature gauge theory.  

A semiclassical model for non-zero temperature QCD has first been
formulated in \cite{Gross:1981br} starting from Harrington-Shepard (HS)
caloron solutions \cite{Harrington:1978ve} which represent infinite chains of 
time-periodic instanton solutions in $\mathbb{R}^4$. Restricted to the strip 
$S^1 \times \mathbb{R}^3$ HS calorons have an integer-valued 
topological charge $Q_\tx{t} \in \mathbb{Z}$ like BPST instantons. For 
$SU(N)$  HS calorons are realized as $SU(2)$ embeddings. 
More restrictive from our point of view is the fact that their vector 
potentials behave asymptotically such that trivial holonomy emerges, 
i.e. their asymptotic Polyakov loop values belong to the center of the group
\beq
\Hol \equiv \lim_{|\vec{x}| \to \infty} \Pol(\vec{x}) \in Z(N)\,.
\eeq
If one would have to describe only the deconfinement phase, 
this asymptotic behavior which is directly related
to the broken center symmetry values of the order parameter 
is certainly adequate. Is this the case also for the confinement phase ?
The authors of \cite{Gross:1981br} have argued that the 
path integral evaluated in the infinite volume limit cannot be dominated by 
classical fields having a non-trivial holonomy. But this conclusion does 
not hold for a single topological excitation created in a finite box or  
in the case of a finite density of such objects.
Therefore, the question arises, whether there are also classical fields 
asymptotically matching to a vanishing order parameter as required in the 
confinement phase. 
Indeed, selfdual time-periodic caloron solutions of the $SU(N)$ Yang-Mills 
equations with arbitrary (non-trivial) holonomy have been explicitly 
constructed by Kraan and van Baal \cite{Kraan:1998pm}, and Lee and Lu 
\cite{Lee:1998bb} several years ago. The {\it new} $SU(N)$ calorons -- 
we call them KvB calorons -- with integer-valued topological charge 
$~Q_\tx{t}~$ consist of $~N~|Q_\tx{t}|~$ constituents, which 
appear as static BPS monopoles \cite{Prasad:1975kr,Bogomolnyi}
if sufficiently separated. Their masses $8\pi^2~T~\nu_m$ 
are then determined by the 
differences of holonomy eigenvalues $\nu_m=\mu_{m+1}-\mu_m\,,$
\beq
\Hol = g\;\exp(2\pi
i\;\diag(\mu_1,\mu_2,\ldots,\mu_N))\;g^\dagger
\eeq 
ordered such that $~\mu_1<\mu_2< \cdots <\mu_{N+1}\equiv 1+\mu_1~$ and 
$~\mu_1+\mu_2+\cdots+\mu_N=0$. 
The separated and massive caloron constituents
can be localized by one or more of the following
criteria \cite{VanBaal:2001pm}:
\begin{enumerate}
\renewcommand{\theenumi}{(\alph{enumi})}
\renewcommand{\labelenumi}{\theenumi}
\item at the centers of their three-dimensional spherical lumps of action 
  or topological charge,
\item at the positions, where the zero-mode density of the Dirac operator
  in the background of the caloron field
  $\slashed{D}=\gamma^\mu(\partial_\mu+A_\mu^{\mathrm{caloron}})$ is maximal, 
  depending on the temporal boundary condition for the fermion field,
\item at the points $\vec{x}$, where at least two of the eigenvalues of
      $\Pol(\vec{x})$ coincide, 
\item at the positions of static Abelian (anti)monopoles representing 
      defects of an appropriately chosen Abelian gauge.
\end{enumerate}
However, if the constituents are close to each other or if they become even
massless, these criteria do not hold unambiguously. The first two criteria
cannot distinguish between the constituents at all. 
The constituents form joint non-static caloron lumps of topological charge. 
The last two criteria are still working, but the extracted positions will 
not completely agree. 

In any case, the non-trivial holonomy suggests a proper
link between the classical background field and the dynamics of the order 
parameter. Therefore, a semiclassical (variational) approach to Yang-Mills 
theory should be based on the new, KvB calorons.
That such an approach is promising has been recently argued by Diakonov
\cite{Diakonov:2002fq}. He and his collaborators have shown that the quantum 
amplitude in the background of KvB calorons leads to interactions of the 
monopole constituents at the quantum level 
\cite{Diakonov:2004jn,Diakonov:2004qs}. 
These interactions become attractive at higher temperature (i.e. deconfinement) 
and repulsive at lower temperature (confinement). Thus, in the deconfinement 
phase HS calorons might be favored (with all but one constituent massless, 
{\it i.e.} invisible), whereas in the 
confinement case the repulsive interactions support a view of a gas to be 
described fully in terms of monopoles or KvB caloron constituents. 
The calculation of the fluctuation determinant still awaits extension from the 
$SU(2)$ case~\cite{Diakonov:2004jn,Diakonov:2004qs} to arbitrary
$N$. The zero-mode contribution to the semiclassical integration measure can 
be obtained for arbitrary $N$ from the determinant of the explicit moduli space 
metric as computed by Kraan a few years ago~\cite{Kraan:1998pn}. 
This, as well as a more detailed analysis of how in particular the parameters 
of the $N$ monopole constituents are related to the scale parameter of the 
standard $SU(N)$ instanton, can be found in Ref.~\cite{Diakonov:2005qa}.

In recent papers \cite{Ilgenfritz:2002qs,Ilgenfritz:2004ws} published 
together with Martemyanov, Shcheredin and Veselov we have explored 
the existence of KvB calorons in the (simpler) case of 
$SU(2)$ lattice gauge theory. We first relied on the {\it cooling} method 
in order to search for (approximate) solutions of the Yang-Mills equations of 
motion. Later on we have investigated smoothed configurations obtained by
four-dimensional smearing i.e. ``keeping closer'' to equilibrium fields 
\cite{Ilgenfritz:2004zz}. In both cases, starting from Monte Carlo equilibrium 
fields generated within the confinement phase we have shown that non-trivial 
holonomy with $\tr\Pol_{\infty} \simeq 0~$ dominates 
the vacuum structure. The topological lumps we have found were
typically composed of monopole constituents and possessed all properties of 
KvB solutions (or their constituents).  

In this paper we are extending our investigations to the $SU(3)$ case of 
finite temperature lattice gauge fields in the confinement 
phase. Relying on the cooling procedure our aim is again to explore
the structure of (semi)classical fields at the level of a few instanton 
actions $~S = n~S_\tx{inst}, ~n=1, \ldots, 4$ with 
$S_\tx{inst}=8 \pi^2/g^2$. For this we had to refine
our methods and to improve lattice observables.  
Our paper is organized as follows. In \sref{analytical} we 
review known analytical formulae for the $SU(3)$ KvB caloron.%
\footnote{For further reading about calorons consider the 
literature, e.g. \cite{Bruckmann:2002vy,Braam:1989qk,Nye:2001hf,Ward:2003sx}.}
Later on, in \sref{latticiser}, we describe the observables and the
methods which are used to compare the analytical results with the classical 
lattice gauge fields. Results for typical examples of cooled lowest-action
(classical) fields are presented in \sref{latticecalorons}. 
The results obtained for ensembles of classical fields extracted also at 
higher action level are discussed in \sref{calensembles}. 

\section{Properties of KvB calorons}\label{analytical} 
\begin{figure}
\vspace{-0.8\fs}
\includegraphics[height=3\fs]{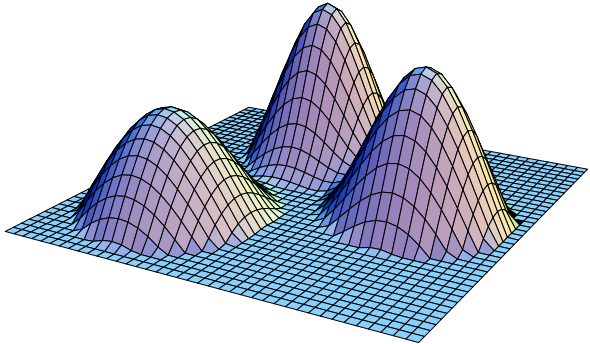}
\vspace{-0.8\fs}
\caption{Logarithmic plot of the action density of a $SU(3)$ caloron 
(from \cite{vanBaal:1999bz}).}
\label{fig:action1}
\end{figure}

Most analytical efforts are currently directed to improving the knowledge of
higher charge $SU(2)$ calorons \cite{Bruckmann:2004nu}. Expressions are known 
for topological charge-one calorons in $SU(2)$ \cite{Kraan:1998xe} 
and for generic $SU(N)$ \cite{Kraan:1998sn}. 
We recall the expressions of the action $~s(x)~$ and the density of the fermionic 
zero-mode $~\Psi_z^\dagger\Psi_z(x)~$ of the $SU(3)$ single-caloron field 
$~A_\mu~$ with
$~|Q_\tx{t}|=1~$. The zero-mode depends on the temporal boundary condition
imposed, $~\Psi_z(x_0+b,\vec{x})=e^{-2\pi iz}\Psi_z(x_0,\vec{x})$ (with $b=1/T$). 
 \bea
 \label{eqn:action}s(x)&=&-\half\,\partial^2\partial^2\ln\psi(x_0,\vec{x})\\
 \label{eqn:fermi}\Psi^\dagger_z\Psi_z(x)&=&-\frac{1}{4\pi^2}\;
 \partial^2\hat{f}_x(z,z)\\
 \label{eqn:gfield}A_\mu(x)&=&\half\,\phi^\half{C_\mu}\phi^\half+
 \half\,[\phi^{-\half},\partial_\mu\phi^\half] 
 \eea
The auxiliary fields in {Eqs.~(\ref{eqn:action}--\ref{eqn:gfield})}
are defined in the Appendix. 
The caloron moduli space consists of three (constituent) monopole
positions $~\vec{y}_m~$ in $~\mathbb{R}^3$ ($m=1,2,3$ and
$~\vec{y}_4\equiv\vec{y}_1$) 
and three phases in $U(1) \times U(1)$ of which one plays the role 
of the overall $S^1$ time location. The temporal extension is scaled to 
$~b=1~$ and the holonomy is left arbitrary.
For illustration the action density for a $SU(3)$ KvB solution with
well-dissociated monopole constituents is plotted in
\fref{fig:action1}.

\begin{figure*}
\centering 
\includegraphics[height=1.2\fs]{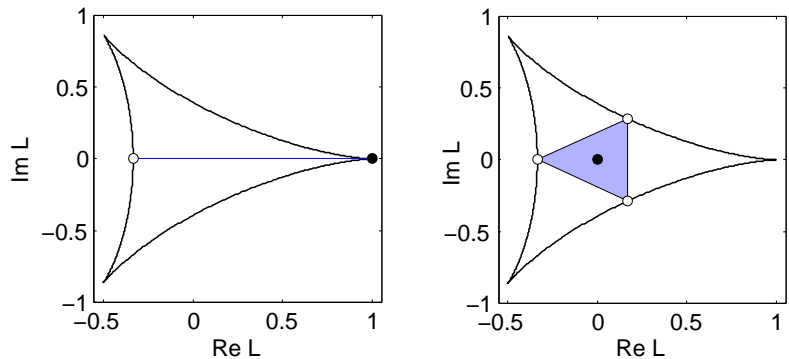}
\caption{
Schematic scatterplots (left: on the real axis, right: inside the triangle) 
for the local Polyakov loop values $~L(\vec{x})~$ in the complex plane, 
for a caloron with trivial holonomy $~L_{\infty} = +1~$ (left) and with
maximally non-trivial holonomy $~L_{\infty}=0~$ (right). Black dots indicate 
the holonomy values, whereas white circles on the curved boundary 
give the Polyakov loop values at the monopole constituents according to 
\mref{eqn:polinterplay} where two eigenvalues of $~\Pol~$ become degenerate 
in accordance with the monopole criterion (c) in \sref{intro}.
} 
\label{fig:weylplot}
\end{figure*}
Provided the constituents are well separated, the Polyakov loop values
at their positions $~\vec{y}_m,~m=1,2,3~$ are 
\bea
\nonumber
\Pol(\vec{y}_1)&=
&\diag(\hphantom{-}e^{-\pi{i}\mu_3},\hphantom{-}e^{-\pi{i}\mu_3},\hphantom{-}
e^{2\pi{i}\mu_3}),\\  \label{eqn:polinterplay} 
\Pol(\vec{y}_2)&=
&\diag(\hphantom{-}e^{2\pi{i}\mu_1},\hphantom{-}e^{-\pi{i}\mu_1},\hphantom{-}
e^{-\pi{i}\mu_1}),\\
\nonumber
\Pol(\vec{y}_3)&=
&\diag(-e^{-\pi{i}\mu_2},\hphantom{-}e^{2\pi{i}\mu_2},-e^{-\pi{i}\mu_2}).
\eea
%
%
In \fref{fig:weylplot} we illustrate the range of Polyakov loop values
$~L(\vec{x}) \equiv (1/3) ~\tr\Pol(\vec{x})~$ for two limiting 
cases: on the left hand side for a $SU(3)$ KvB caloron with trivial holonomy 
$~L_{\infty} \equiv (1/3) ~\tr\Pol_{\infty} = +1~$
which corresponds to a $SU(2)$ embedding of a Harrington-Shepard caloron 
\cite{Harrington:1978ve} with real values of $~L(\vec{x})~$ only, and on 
the right hand side for maximally non-trivial holonomy $~L_{\infty} = 0~$ 
leading to dissociated equal-mass constituents and $~L(\vec{x})~$ populating
the inner triangle. 

The features of KvB calorons are the following. 
\begin{enumerate}
\item The holonomy can be chosen arbitrarily. 
\item The multi-caloron contains $N~|Q_\tx{t}|$ constituent monopoles in
  accordance with various definitions.
\item  The fermionic zero-mode density $\Psi_z^\dagger\Psi_z(x)$ is localized 
  at the $m^\tx{th}$ constituent if $z\in(\mu_m,\mu_{m+1})$ and jumps
  (delocalizes) for $z=\mu_{m}$. 
\end{enumerate}
These remarkable features will be illustrated for calorons of various
topological charge $Q_\tx{t}$, which are obtained on asymmetric lattices,
i.e. $N_\tau \times N_\sigma^3$ with $N_\sigma>N_\tau$, by the cooling
technique. This should be seen as an illustration of an extended set of 
caloron solutions. 

If realized on the lattice, the holonomy obviously cannot be chosen freely
but dependent on the actual finite-temperature phase.
Inspired by the properties listed above,
Gattringer and Schaefer~\cite{Gattringer:2002wh,Gattringer:2002tg} have 
analyzed a subsample of 
$Q_\tx{t}=\pm 1$ Monte Carlo configurations, taken from the confinement 
and deconfinement phase, for the localization of the (single) zero-mode 
with changing $~z~$. In the confinement phase, identified with maximally
non-trivial holonomy, they found the pattern of zero-mode jumping. In the 
deconfined phase, they found a wide range in $~z~$ where the zero-mode 
didn't change localization, becoming delocalized only at one angle $~z~$
corresponding to the phase of $L_{\infty}$. 
The localization of the zero-mode wave function was in agreement
with local clusters of equal sign of the topological charge 
density~\cite{Gattringer:2003uq}. Globally, however, the latter was far 
more complex than that of a single caloron. 

Notwithstanding this state of affairs it is suggestive to carry over the 
lattice observables developed for the purpose of analyzing classical
configurations to a program to rediscover superpositions
of KvB calorons and anticalorons in Monte Carlo ensembles. 
Our aim is eventually to prove the validity of the semiclassical approach
near to the transition temperature and above. 

\section{Lattice tools of the analysis}\label{latticiser}
Working on the lattice has two obvious disadvantages -- the explicit
breaking of scale-invariance and the existence of lattice artifacts. 
The classical action of a multi-caloron or instanton solution with topological 
charge $~Q_\tx{t}~$ is $~S = |Q_\tx{t}|~S_\tx{inst}$.
It does not depend on the parameters of the solution: group orientations, 
positions and scale parameters. On the contrary, the standard Wilson action
\beq
S^W = \sum_{x,\mu<\nu} (1 - \frac{1}{3}~\Re~\tr~U_{x,\mu\nu}),
\eeq
depends on the scale size $~\rho~$ of an instanton discretized on the lattice.
It can be calculated as \cite{GarciaPerez:1994ki,Bruckmann:2004ib}
\beq
S^W_\tx{inst}=S_\tx{inst} \left(1+d_2\!\left(\frac{a}{\rho}\right)^2+
\mc{O}(a/\rho)^4\right)
\eeq
with $d_2=-1/5$. Hence this lattice action favors small instantons. For a KvB caloron, the
intrinsic scale is determined by the constituent separations expressed by 
$~\rho_m~$, and similarly small $~\rho_m~$ are preferred. 
For $SU(2)$, in the limit of separated constituents, $~\rho_m = d~b/\pi^2~$
with $~b=1/T~$ and $~d~$ being the distance between the two 
constituents. As a remedy one could choose an (improved) action 
with zero or even positive coefficient $d_2$. For the purposes of this 
paper, however, we preferred to keep the 
numerically cheaper Wilson gauge action as the cooling action and 
simply to monitor the scale-size dependence.

Due to the finiteness of the lattice one cannot reproduce the
$S^1\times\mathbb{R}^3$ infinite toroidal or cylindric topology. 
Instead, depending on the aspect ratio $N_\tau/N_\sigma$ one can 
interpolate between the topologies of the 4-torus $T^4$ and that of 
a cylinder. On the other hand, using the lattice as stage of such 
investigations has also advantages, for example the possibilities 
\begin{itemize}
\item[\ldots]  to produce classical fields in different topological sectors 
           without knowing explicitly the corresponding analytic expressions,
\item[\ldots]  to include fermions into the study of solutions of 
           the pure gauge theory,
\item[\ldots]  to test the predictions of corresponding semiclassical models 
           and 
\item[\ldots]  to explore the semiclassical structure of the 
               respective quantized gauge field theory 
               (also with dynamical fermions)   
\end{itemize}
within one and the same framework.

In order to generate a large number of typical classical solutions of the 
Euclidean $SU(3)$ Yang-Mills equations of motion for a given temperature
and volume, Monte Carlo lattice gauge fields have been cooled with respect 
to the standard Wilson gauge action. The smoothed configurations have then be 
characterized with the help of an $\mc{O}(a^4)$ improved field strength 
tensor \cite{deForcrand:1997sq} 

\beq
F_{\mu\nu}(x)=\frac{1}{4}\!\!\sum_{x\in\tx{loops}}
  \left[\frac{3}{2}\,W^{1,1}_{\mu\nu} 
- \frac{3}{20}\,W^{2,2}_{\mu\nu}
+ \frac{1}{90}\,W^{3,3}_{\mu\nu}\right],
\label{eqn:fieldstrength}
\eeq

\noi
where $~W^{k,l}~$ is the traceless and antihermitean part of the
respective $k \times l$ rectangular Wilson loop. 
The sum is extended over the four loops of 
quadratic shape and each given size $~(n,n)~$ in the $\mu\nu$-plane
which have the site $x$ as a common corner.
From the field strength one can evaluate the corresponding improved 
operators of action density $s(x)$ and topological charge density 
$q_\tx{t}(x)$ 
\bea
s(x)&=&\frac{1}{2g^2}\;\tr(F_{\mu\nu}^{\,2}(x)) \quad\tx{and}\\
q_\tx{t}(x)&=&-\frac{1}{32\pi^2}\,\epsilon_{\mu\nu\rho\sigma}\,
\tr(F_{\mu\nu}(x)F_{\rho\sigma}(x)).
\eea 
In our case of cooled gauge fields, the total topological charge 
$Q_\tx{t}=\sum_x q_\tx{t}(x)$ deviates from integer values typically by 
less than one percent.

The starting configurations for our cooling search have been produced 
by Monte Carlo simulation at $~\beta \equiv 6/g^2 = 5.65 \,.$ 
For $N_t = 4$ and $~6$  the Monte Carlo samples 
are in the confinement phase with $~T/T_\tx{dec} \approx 0.91$ and $0.6$, 
respectively. In these cases the volume average of the Polyakov loop vanishes. 
Under the influence of cooling this average starts to diffuse.  
A large fraction of configurations will approximately remain in the vicinity of 
maximally non-trivial holonomy until each turns into a classical solution.
Various lattice sizes $~4 \times 12^3$, $6 \times 12^3$, $4 \times 20^3$, 
$6 \times 20^3$, and $12^4~$ have been chosen in order to study how the yield 
of classical configurations (populating different parts of the caloron
parameter space) depends on the lattice size, in particular on the aspect
ratio. Starting from the deconfined phase ($T>T_\tx{dec}$), mainly fields with 
trivial holonomy have been obtained by cooling. Throughout this 
paper we restrict ourselves to the confinement phase since we are mainly 
interested in KvB solutions with generically non-trivial holonomy.  

To what extent a lattice gauge field can be regarded to be classical is
measured by the violation of (anti) selfduality $\delta_F$, defined
by 
\beq
\delta_F ~=~
\sum_x ~\left|\frac{g^2}{8\pi^2}s(x)-\left|q_\tx{t}(x)\right| \right|.
\eeq
For a strictly selfdual or anti-selfdual gauge field this quantity
vanishes. 

If the caloron constituents are well-separated the action density of
the caloron becomes static in $x_0$. An auxiliary observable which we call
{\it non-staticity} 
\beq
\delta_t ~=~ \frac{N_t}{4}~
\frac{\sum_{x}{\left|s(x+\hat{0})-s(x)\right|}}{\sum_x s(x)}
\eeq
has been introduced to distinguish between dissociated 
monopole constituents (for which the action density is static) 
on one hand and the non-dissociated case (which gives rise to normal
time-dependent instanton lumps) on the other.
The factor $N_t/4$ compensates for finite discretization 
as was chosen in previous work for $SU(2)$ \cite{Ilgenfritz:2004ws}.  
For classical solutions, evaluating
the continuum action density \mref{eqn:action} on a grid of points, 
the following benchmarks have been calculated for a symmetric set-up of the
constituents. Assuming that the holonomy is maximally non-trivial,
i.e. $\tr(\Hol)=0$, at least two constituents are discernible as 
maxima of the action density if $~\delta_t~$ is lower than some
bifurcation value $~\delta_t^*=0.27$. If the non-staticity is even
smaller than $~\delta_t^{**}=0.06~$, then it is possible to resolve all 
three constituents in the analytically known action density profile of the 
$SU(3)$ KvB caloron~\cite{BMprivate}.  


Unlimited cooling with the standard 
Wilson action will always drive the field towards 
a topological trivial vacuum $|Q_\tx{t}| \to 0$. In order not to loose all 
approximately classical fields the cooling process is monitored and a 
stopping condition is chosen. Using the non-staticity 
$\delta_t$ and the violation of (anti) selfduality $\delta_F$ we have
employed two different stopping criteria which allowed us to catch an 
abundance of (multi-)caloron solutions in order to explore the moduli 
space of calorons. Cooling is stopped,    
\begin{enumerate}
\renewcommand{\theenumi}{(\Alph{enumi})}
\renewcommand{\labelenumi}{\theenumi}
\item\label{stop1} if the violation of (anti)
  selfduality $\delta_F$ passes through a minimum 
  with the constraint $\delta_F<0.1$,%
  \footnote{This criterion is found to be practically equivalent to 
  a stopping condition used in our previous 
  papers \cite{Ilgenfritz:2004ws,Ilgenfritz:2002qs}.}  
\item\label{stop2} if either $\delta_F$ or the non-staticity $\delta_t$ 
  passes through a minimum, depending on what happens first. Additionally 
  the (less restrictive) 
  constraint $\delta_F<0.2$ is applied.
\end{enumerate}%
\noindent
Both criteria result in nearly (anti) selfdual gauge fields
representing various topological sectors. In particular, in most cases
configurations with some integer topological charge $Q_\tx{t}$ and action 
$S \gtrapprox S_\tx{inst}~|Q_\tx{t}|$ (implying a slight violation $\delta_F$) 
are found.


With the traditional choice \ref{stop1} gauge fields are cooled until
the equations of motion are fulfilled as good as possible and only
violated to a small extent. There are some reasons to argue that this
stopping condition is too strict. Superpositions of semiclassical
objects (e.g. instanton and anti-instanton) are no strict solutions of
the equations of motion. Nevertheless, a mixture of both must exist in a 
realistic model of the vacuum structure.  On the other hand it is impossible 
to model the moduli space for an infinite $\mathbb{R}^3$ volume. The numerical
investigation can only be done for a discrete torus, where  it is
known that a $|Q_\tx{t}|=1$ field can exist only at the expense of violating
the (anti) selfduality \cite{Braam:1989qk}. Accepting a certain
violation of the equation of motion constitutes a compromise between
the moduli spaces of the torus and the finite temperature space-time
$S^1\times\mathbb{R}^3$. Furthermore lattice artifacts of the action
will bias the semiclassical content of fields. Hence one should not
insist on perfect solutions of the (lattice) equations of motion. Stopping
condition \ref{stop2} allows to stop in an earlier stage. In particular this 
results in an enrichment of static, approximate classical configurations. 


For a cooled gauge field configuration we determine the 
holonomy $\Hol$ as an average over a spatial region $V_{\alpha}$, 
where the action density is minimal,  
\bea
\nn\Hol&=&\frac{1}{V_{\alpha}}\sum_{\vec{x} \in V_{\alpha}}
\left[\Pol(\vec{x})\right]_\tx{diagonal} \\ &\equiv&
\frac{1}{V_{\alpha}}\sum_{\vec{x} \in V_{\alpha}}
\diag(e^{i\phi_1(\vec{x})},e^{i\phi_2(\vec{x})},e^{i\phi_3(\vec{x})})
\eea
with
\beq
\Pol(\vec{x})=\prod_{x_0=1}^{N_\tau}U_{(x_0,\vec{x}),0} \,.
\eeq
The domain $V_{\alpha}$ is taken as the fraction of
$\alpha=10\,\%$ of the 3-dimensional volume for which the spatial action density
$\bar{s}(\vec{x})$ (the $x_0$-average of $s(x)$ with fixed $\vec{x}$) is
smaller than everywhere in the remaining region. To get a meaningful result
for the $\Hol$ operator it is necessary to choose a
suitable gauge. We decided to take the Polyakov loop diagonalizing 
gauge, {\it i.e.} we average over the eigenvalues of $\Pol$.    
From the observed Polyakov loop operator $\Pol(\vec{x})$ not only the 
holonomy $\Hol$ (and the $\mu_m$'s) is determined, but also the 
(monopole) positions $\vec{y}_n$ and consequently the overall number 
of constituent monopoles.   

Indeed, with the diagonalized Polyakov loop operator
$~\left[\Pol(\vec{x})\right]_\tx{diagonal}~$
we identify the constituents by  determining the positions $\vec{Y}$
where two eigenvalues approach each other. For the
caloron solution in the continuum they would exactly coincide. On the
lattice this condition has to be slightly relaxed. Thus, in our numerical
work we fixed monopole positions $\vec{Y}$ 
through the local minima of the function    
\beq
f(\vec{x}) = \min_{i,j}|e^{i\phi_i(\vec{x})}-e^{i\phi_j(\vec{x})}|
\label{eqn:monopolefinder}
\eeq
in the 3-dimensional volume with the additional constraint that some
$\phi$ exists such that
\beqs
\min_{\phi\in[0,2\pi)}\left|L(\vec{Y})-
(2e^{i\phi}+e^{-2i\phi})/3\,\right| < 0.05.
\eeqs
This ensures that (approaching some $\phi$) the eigenvalues of $\Pol$
are sufficiently degenerate at $\vec{Y}$. To avoid spurious monopole
positions, only monopole constituents with
$~\bar{s}(\vec{Y})>0.1\max_{\vec{x}}\bar{s}(\vec{x})~$ are taken into
account. 

For individual configurations we also compute the spectrum of the
clover-improved Wilson Dirac operator, defined as 
\beas
\slashed{d}(x,y) &=&  \delta_{x,y} - \kappa
\sum_{\mu=1}^{4}\Big[(\unity_4-\gamma_\mu)\;U_{x,\hat{\mu}}\,\delta_{x+\mu,y}
+\\&&\hspace{2cm} (\unity_4+\gamma_\mu)\;U^\dagger_{x-\mu,\hat{\mu}}\,\delta_{x-\mu,y}\Big]\\
\label{eqn:latdirac}
\slashed{D}(x,y)&=&\slashed{d}(x,y)+\frac{i}{2}\,\kappa\,c_\tx{sw}
\,\sigma_{\mu\nu}F_{\mu\nu}(x)\,\delta_{x,y} \\ && \qquad \text{with} \qquad \sigma_{\mu\nu} = \frac{i}{2}\left[ \gamma_\mu,\gamma_\nu
\right].
\eeas
The parameters are chosen to be $\kappa\!=\!1/8$ (massless) and
$c_\tx{sw}\!=\!1$ (tree-level improved). For all spatial directions
the boundary conditions of spinor fields are periodic, for the
temporal
direction one has variable ones: $\Psi_z(x_0+b,\vec{x})=e^{-2\pi iz}\;
\Psi_z(x_0,\vec{x})$. Eigenmodes are computed using the
ARPACK package \cite{ARPACK:package}.
\section{Properties of lattice calorons}\label{latticecalorons}
%
Under the conditions described above, different classical fields with 
$~|Q_\tx{t}|=1,\ldots,7~$ embedded in an arbitrary holonomy are obtained 
from cooling. Since the Monte Carlo simulation is performed in the
confined phase, the yield of (almost) trivial holonomy configurations
is low. Here we present and explain examples for $~|Q_\tx{t}|=0,1,2$. In order
to illustrate these solutions, we provide tables with gluonic and
fermionic observables.
\begin{table}
\centering
\begin{tabular}{|cc|}
\hline
\hline
\multicolumn{2}{|c|}{example 1 -- global observables} \\
\hline
\hline
$S/S_\tx{inst}$      & $\phantom{+}1.14$  \\
$Q_\tx{t}$ & $-1.00$ \\
$\mu_i$      & $\left\{\begin{array}{r}-0.36\\ 0.00\\ 0.36\end{array}\right\}$\\
$n_+-n_-$    & $1-0$\\
$\delta_t$   & $1.3\,\cdot10^{-3}$\\
$\delta_F$   & $0.14$\\
\hline
\hline
\end{tabular}
\caption{Properties of a $Q_\tx{t}=-1$ caloron gauge field 
shown in 
Figs. \ref{fig:q1}, \ref{fig:q1flow}, \ref{fig:q1fit} and \ref{fig:q1abel}.}
\label{tab:gluon1}
\end{table}
  
\begin{figure}[t]
\centering
\includegraphics[height=\fs]{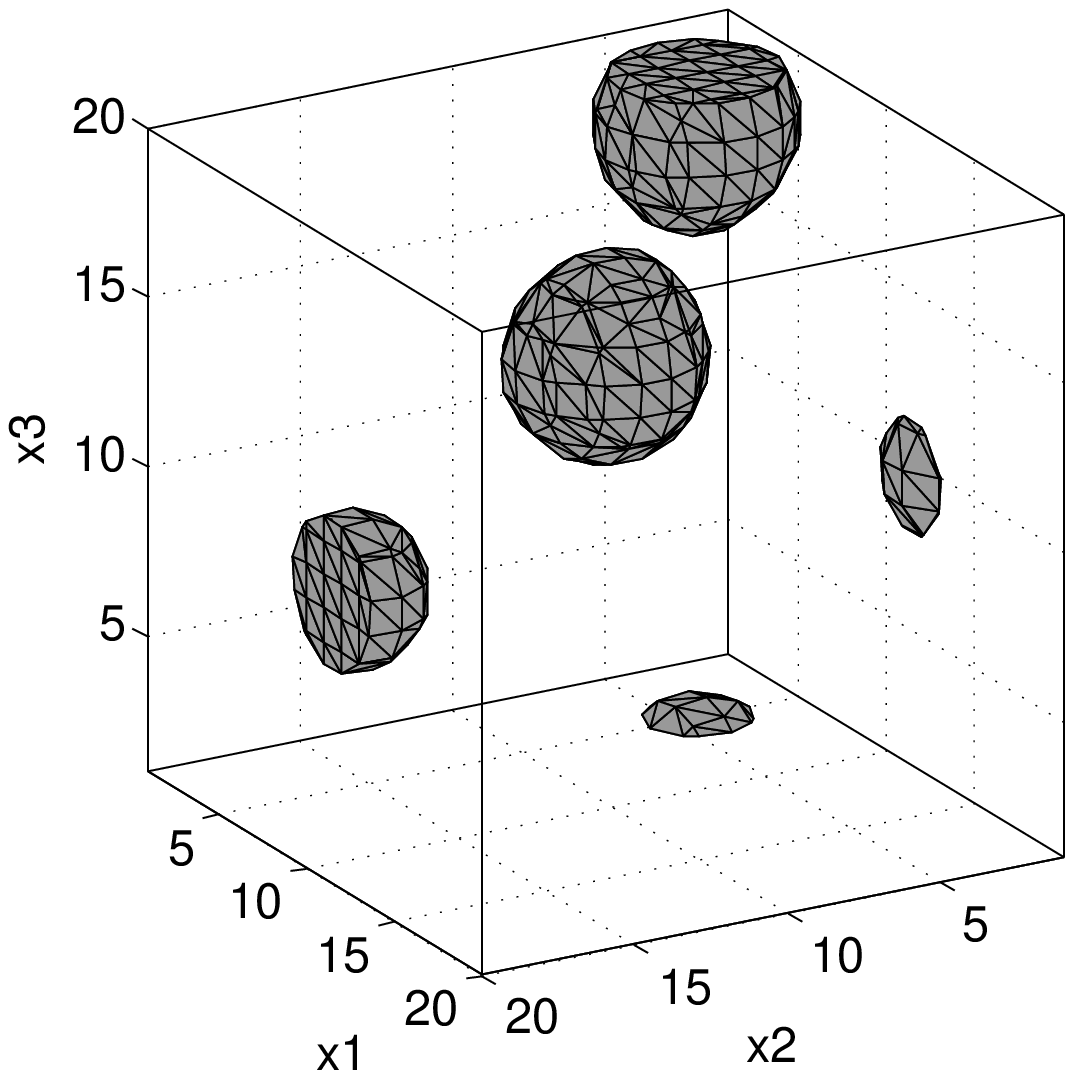}
\includegraphics[height=\fs]{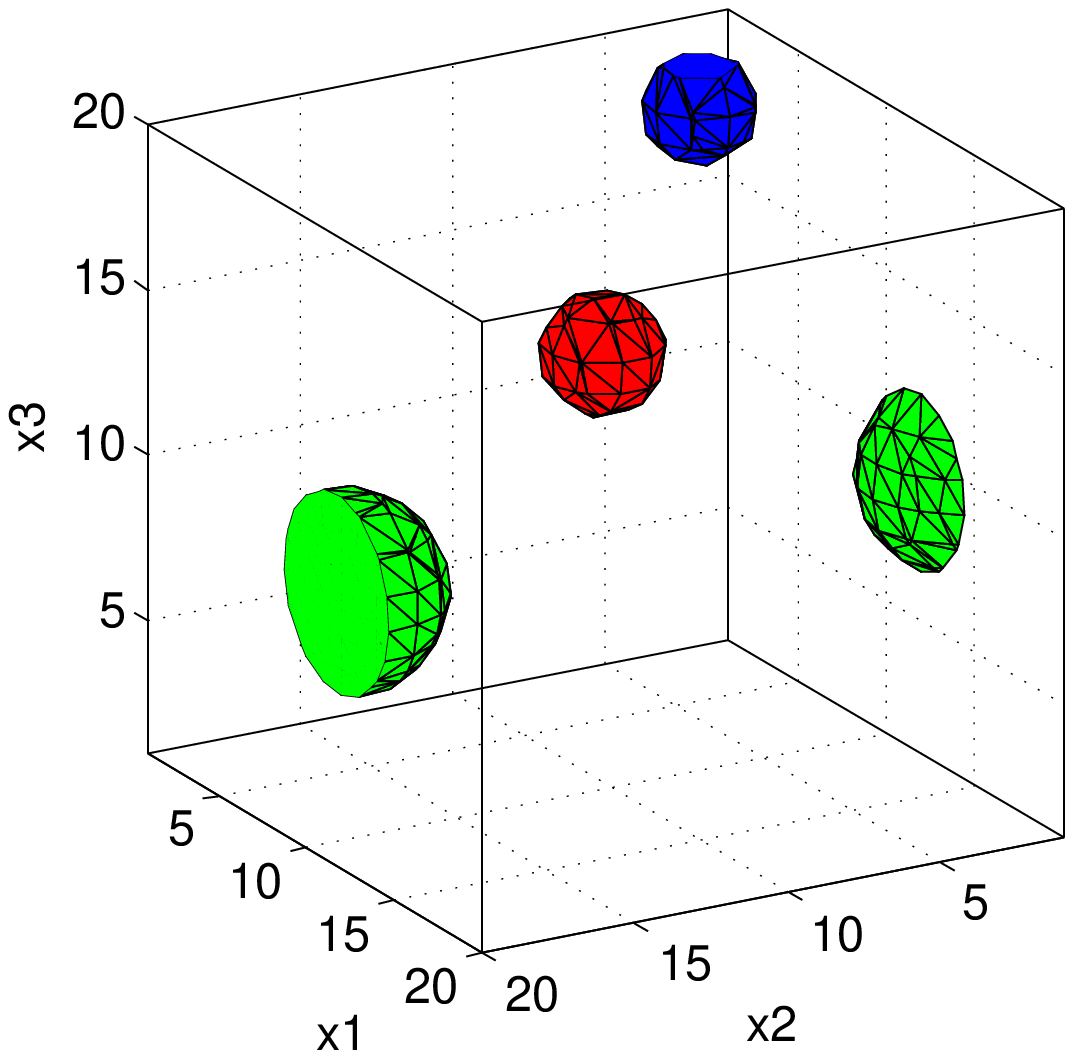}
\vspace{-0.5cm}
\includegraphics[height=\fs]{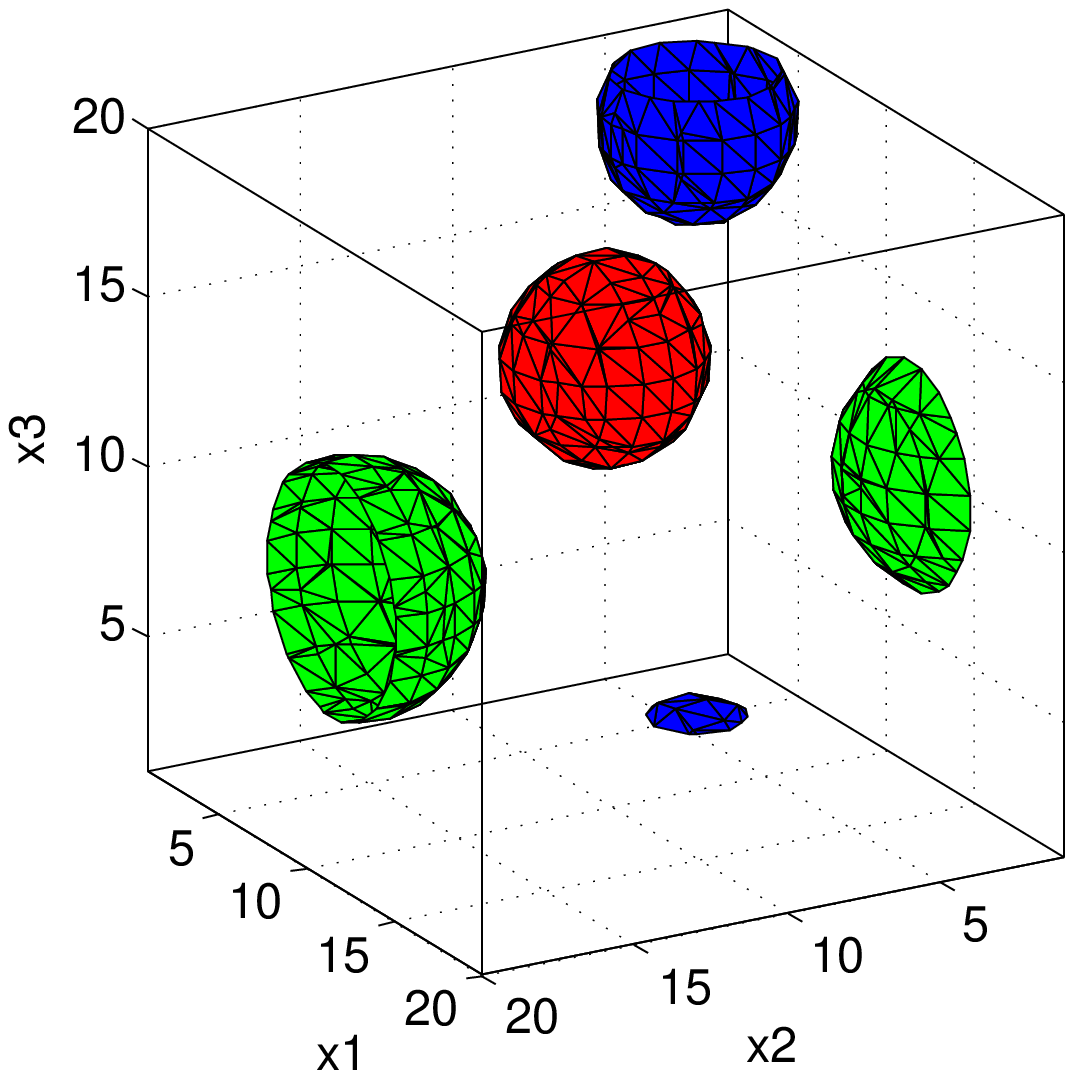}
\includegraphics[height=\fs]{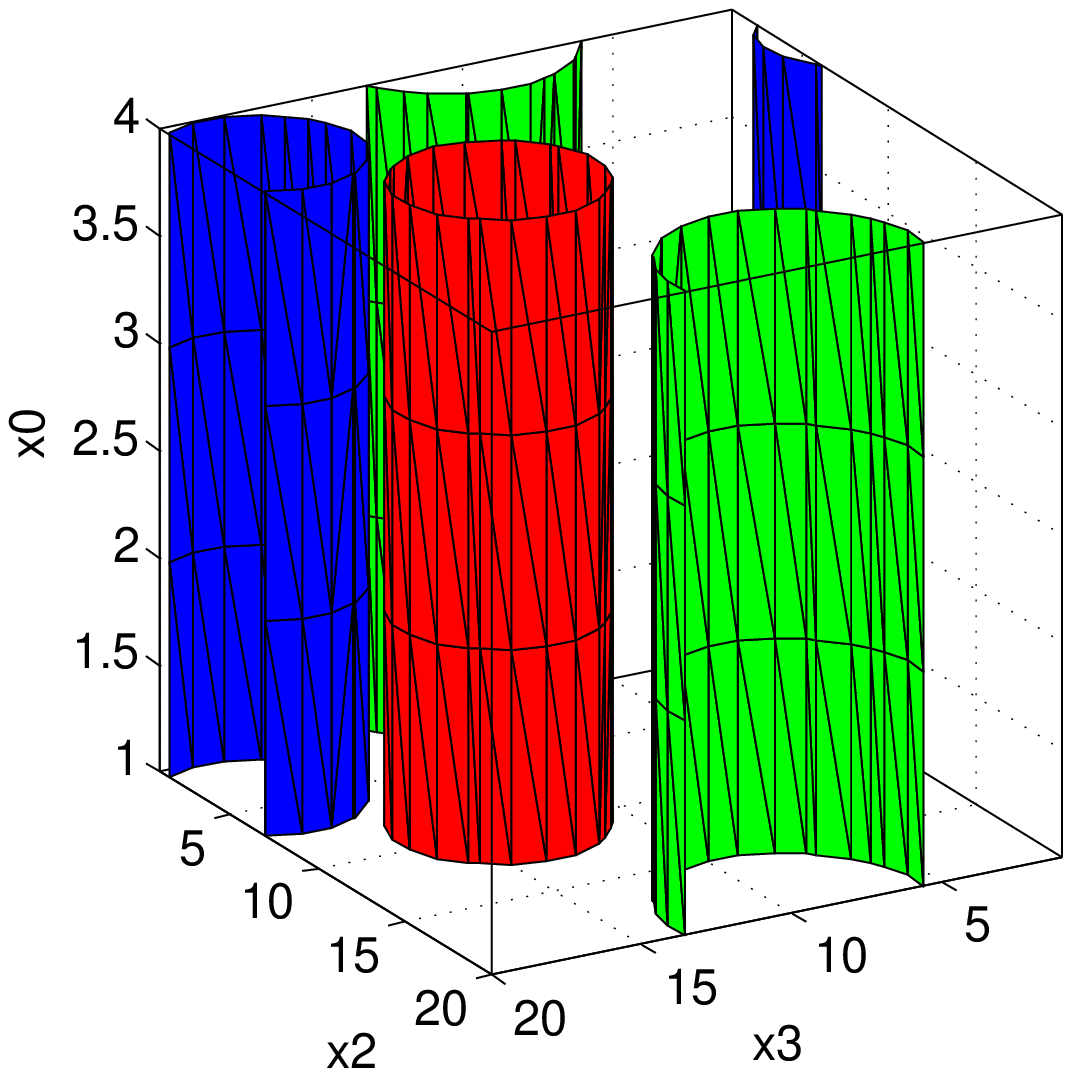}
\caption{Caloron with non-trivial holonomy (example 1), topological charge
$Q_\tx{t}=-1$ and separated
constituents. We show isosurface plots for the action density (u.l.), 
for the function $f(\vec{x})$ (\mref{eqn:monopolefinder}) localizing 
the monopoles, 
for the fermionic density $\Psi_z^\dagger\Psi_z(x)$ with boundary 
conditions $~z=1/6, 1/2, 5/6~$ drawn with different colors/brightness 
at fixed $x_0$ (l.l.) and fixed $x_1$ (l.r.).} 
\label{fig:q1}
\end{figure}

Our first example of a $SU(3)$ classical lattice gauge field has topological
charge $Q_\tx{t}=-1$ and non-trivial holonomy (cf. \tref{tab:gluon1}). It has 
been generated by cooling with respect to the standard Wilson action 
on a lattice of size $4 \times 20^3$ with the stopping criterion
\ref{stop2}. Due to the non-trivial holonomy the gauge field contains
three (almost equal) massive constituent monopoles, which can be
identified by the monopole criteria (a--d). These criteria coincide
since the monopoles are sufficiently separated. In \fref{fig:q1} this is
made visible in the form of isosurfaces of the action density, of the function
$f(\vec{x})$ according to \mref{eqn:monopolefinder} showing
points, where two eigenvalues of  $\Pol(\vec{x})$ come close to each other,
as well as isosurfaces of the fermionic density for varying fermionic 
boundary conditions. 
\noindent
Evidently, the significant features of the KvB caloron of this first 
example are visible in \fref{fig:q1} but the field is not perfectly
anti-selfdual (compare with the $\delta_F$ value in \tref{tab:gluon1}).
This effect is due to the stopping condition \ref{stop2}
which terminated cooling at the minimal value of
$\delta_t$ with the constraint $\delta_F<0.2$. 


Further cooling, until stopping condition \ref{stop1} applies,
would squeeze the constituents together such that no separate constituents 
would be visible in the action or topological charge 
density. Still, the Polyakov loop $\Pol$ and also Abelian monopoles 
would indicate a substructure of the single massive lump. Fermionic observables 
would give only a slight hint to the substructure, since the 
fermionic zero-mode density only slightly moves around
inside the lump and is ``breathing'' under changes of the temporal boundary 
condition of the Dirac operator.
It ``delocalizes'' to an algebraic decay at $z=\mu_i$. 
This situation is not visualized. 

\begin{figure}[t]
\centering
\includegraphics[height=\fs]{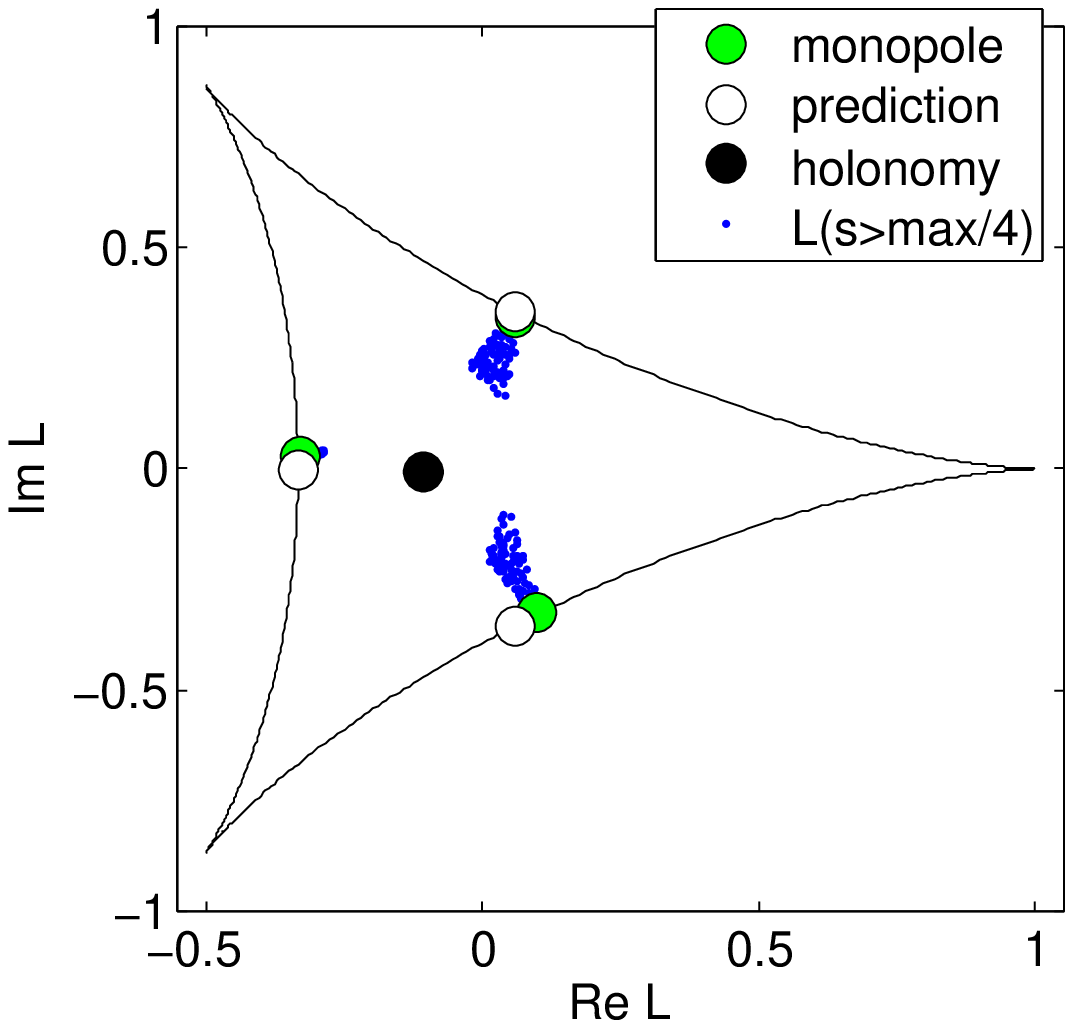}
\includegraphics[height=\fs]{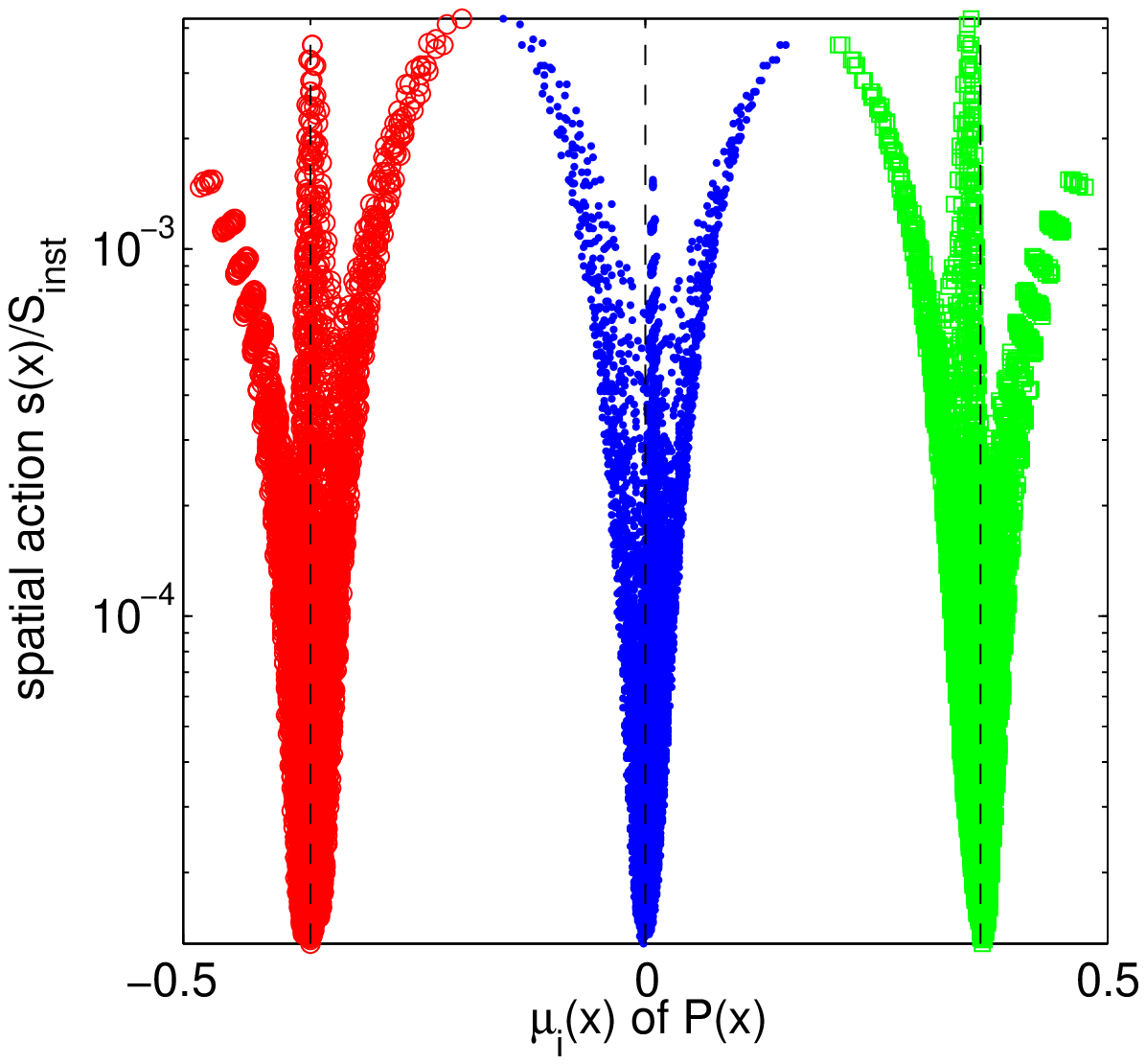}
\includegraphics[height=\fs]{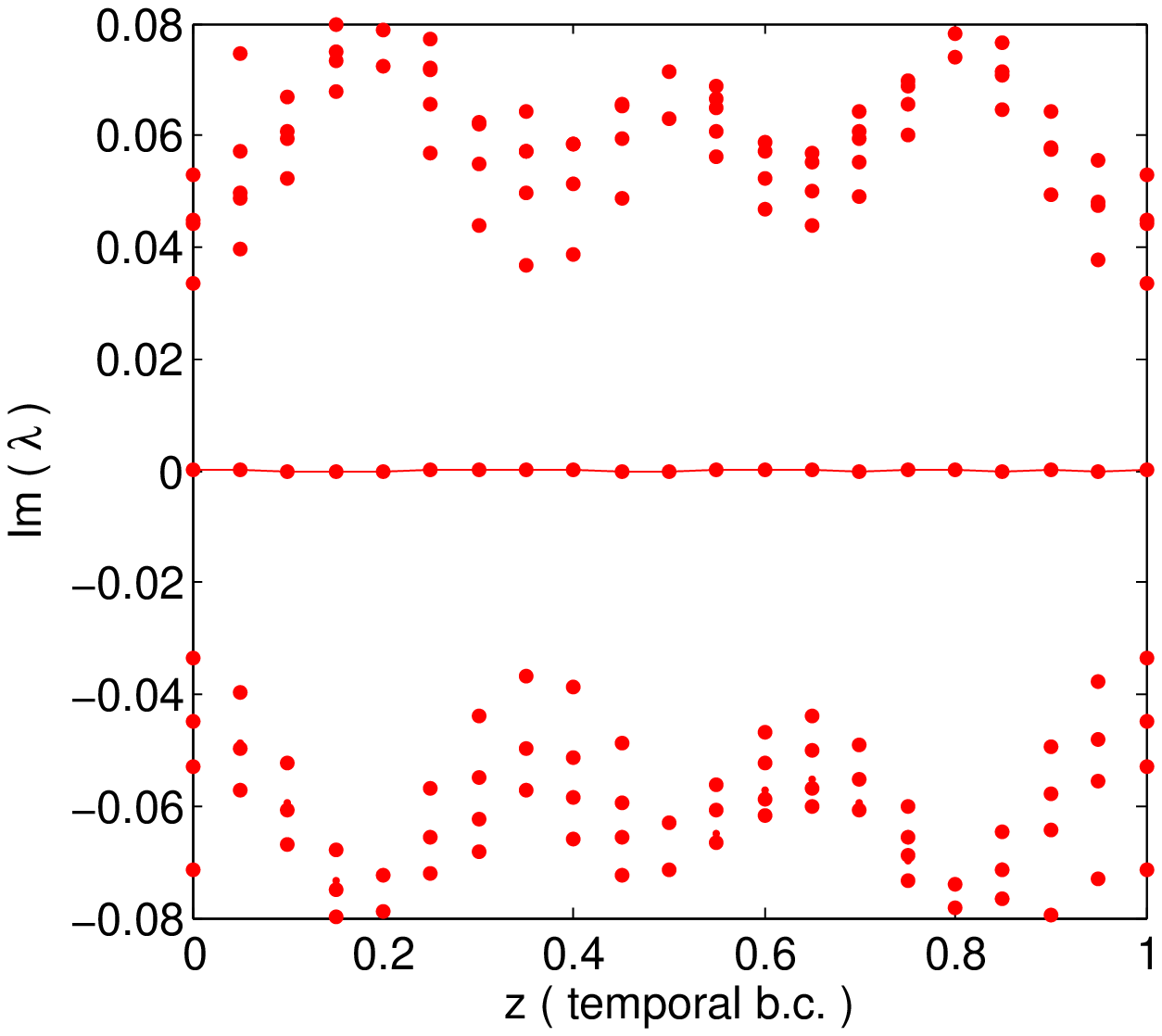}
\includegraphics[height=\fs]{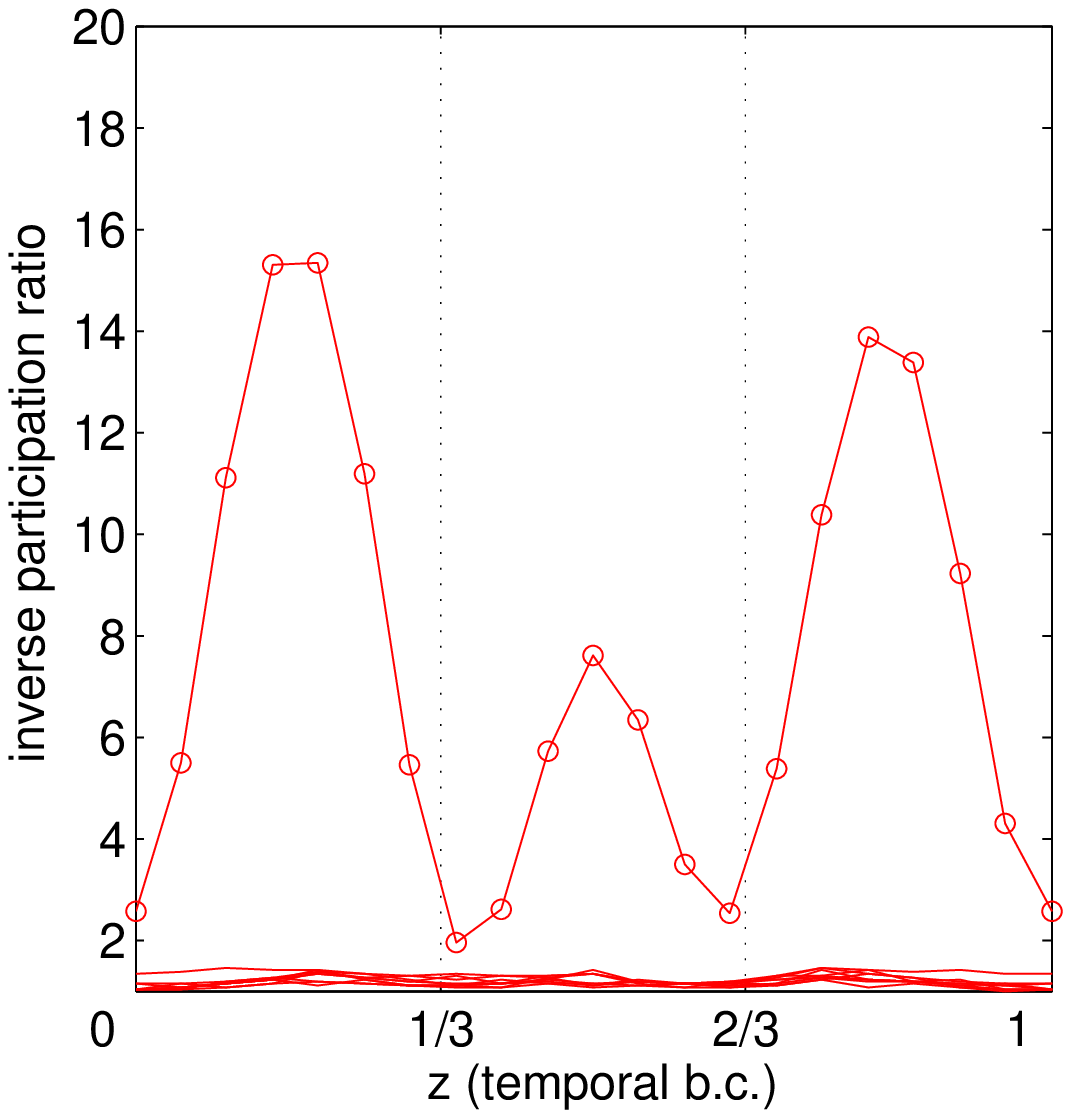}
\caption{Caloron (example 1): scatterplot of $\Pol(\vec{x})$ (for 
  the points with high enough action density) in the
  complex plane (u.l.); eigenvalues of $\Pol(\vec{x})$ versus spatial
  action density (eigenvalues are sorted $-0.5\leq\mu_1<\mu_2<\mu_3\leq 0.5$, 
  as indicated by different points (colors/styles)) (u.r.); $\Im(\lambda)$ 
  (eigenvalue of $\slashed{D}$) (l.l.)
  and inverse participation ratio of the lowest eigenmodes (l.r.) versus
  the temporal fermionic boundary condition $z$.}
\label{fig:q1flow}
\end{figure}

The clover improved Dirac operator with the background caloron field
contains one zero-mode with positive chirality. The localization,
measured by the inverse participation ratio
$~\IPR=V~\sum_x(\Psi_z^\dagger\Psi_z(x))^2~$, strongly depends on
the boundary condition and signals a delocalization $~\IPR\to 1~$ where
the jumping of the zero-mode is expected to occur, namely at
$~z=0;1/3;2/3~$ (cf. \fref{fig:q1flow} lower right). The maximal
localization is expected at $~z=1/6;1/2;5/6$. 

Since the limit $~|\vec{x}|\to\infty~$ is not accessible on a finite lattice 
one might question the possibility to define the holonomy on the
lattice. Nevertheless, as the upper right part of \fref{fig:q1flow}
shows, the eigenvalues of $~\Pol~$ have very small variance over the $3D$
points where the spatial action $~\bar{s}~$ is low.  This property of
the solution makes our definition of $~\Hol~$ {\it a posteriori} plausible. 
Furthermore, the eigenvalues of $~\Pol~$ approach each other at points with 
larger values of the spatial action. Since one does not observe an
exact matching, the monopole criterium relying on
\mref{eqn:monopolefinder} is numerically better defined.
Confirming the analytic expectation, the action density is
static as indicated by the small value of $~\delta_t~$ and resides in
separated constituents.  

\begin{figure}[tbp]
\centering
\includegraphics[height=1.3\fs]{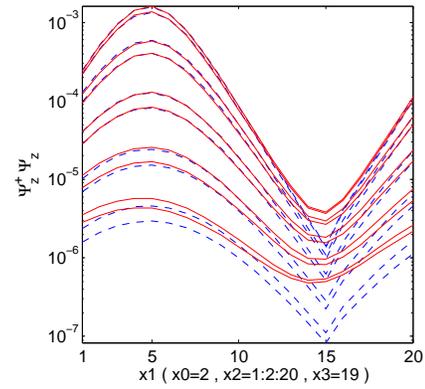}
\caption{Fit of the analytic expression of $~\Psi_z^\dagger\Psi_z(x)~$ to
  the actual density for $~z=5/6~$; full lines means measured; dashed lines
  correspond the fit.}
\label{fig:q1fit}
\end{figure}

\noindent
Fitting the lattice field by analytic expressions of the $SU(3)$ KvB
caloron solution is not feasible. The numerical reconstruction of the
gauge links $~U_{x,\mu}~$ in terms of the vector potential $~A_\mu(x)~$ is
too costly. Moreover, the ``latticized'' analytic solutions are not
yet adapted to the periodic boundary conditions. 
But for gauge invariant quantities like $~s(x)~$ and $~\Psi_z^\dagger\Psi_z(x)~$
the available analytical expressions have been easily used for a fit. 
The zero-mode density is (exponentially) localized 
{\it only at one respective constituent} for almost all angles $z$, 
minimizing in this way effects of the spatial boundary 
condition. This is shown as an example
in \fref{fig:q1fit} and turns out to provide a correct fit of the profile 
over two orders of magnitude.

\noindent

\begin{figure}
\centering
\includegraphics[width=1.5\fs]{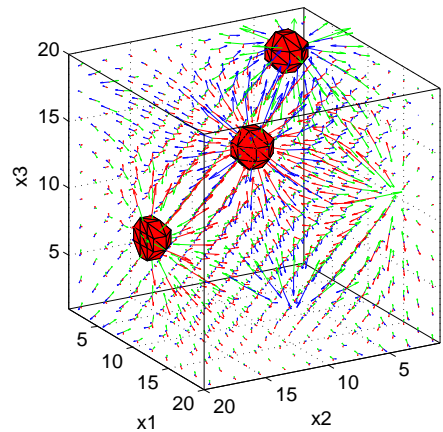}
\caption{Abelian monopoles within a $Q_\tx{t}=-1$ caloron with 
  non-trivial holonomy and separated constituents (example 1); colored arrows
  show electric ($=-$magnetic) field components of Abelian
  projected $U(1) \times U(1)$ field; the isosurface shows the action 
  density of the Abelian projected field.}
\label{fig:q1abel}
\end{figure}

Finally let us have a look at the charge--one caloron in a maximally Abelian
projection which is carried out similar to Ref.~\cite{Langfeld:2003ev}. 
We maximize
$~F_U(D,g)=\Re\;\tr(g(x)\,U_{x,\mu}\,g^\dagger(x+\hat{\mu})\,D_{x,\mu}^\dagger)~$ 
with respect to $~g\in SU(3)$ and $~D~$, where $~D_{x,\mu}~$ is a $U(1)^2
\subset SU(3)$  matrix of the form
$~D=\diag(e^{i\phi_1},e^{i\phi_2},e^{-i(\phi_1+\phi_2)})~$. 
Considering the metric $~|U-D|^2=\tr\,(U-D)^\dagger (U-D)~$ the gauge
transformed link $~U~$ is required to be as close as possible to the diagonal matrix
$~D$. Technically this is performed using the Cabibbo-Marinari method. The static
electric (equal to $~\pm$magnetic part for an (anti)selfdual field)
components of the $U(1)\times U(1)$ field strength are computed 
from the improved field strength tensor \mref{eqn:fieldstrength},
where the $SU(3)$ links $~U_{x,\mu}~$ are replaced by the projected  $~D_{x,\mu}$'s. 
The resulting field pattern is shown in \fref{fig:q1abel}. The maximally
Abelian monopoles (sources of the electric {\it and} magnetic fields)
reside inside the spheres where the Abelian field becomes singular. 

\begin{figure}
\centering
\includegraphics[height=\fs]{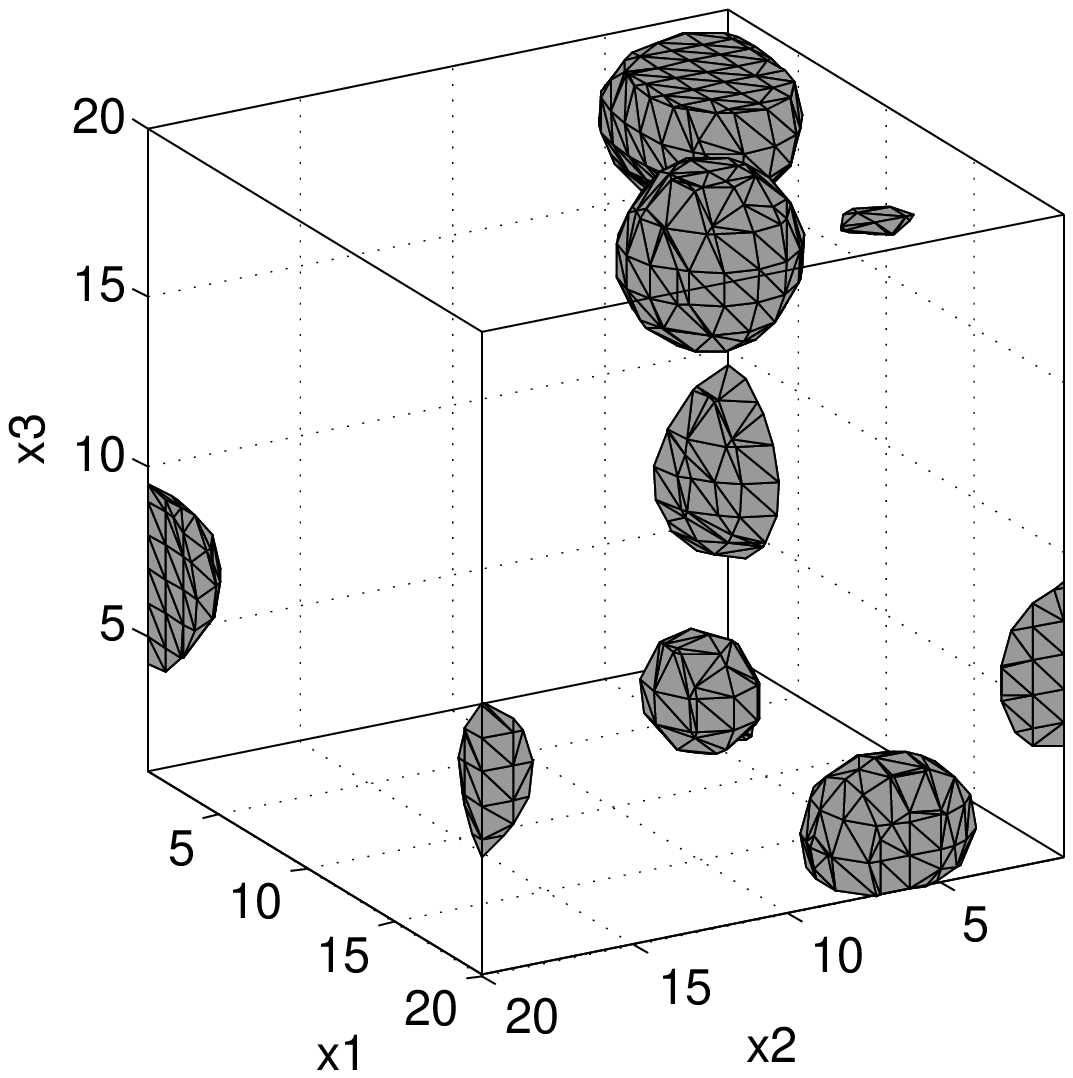}
\includegraphics[height=\fs]{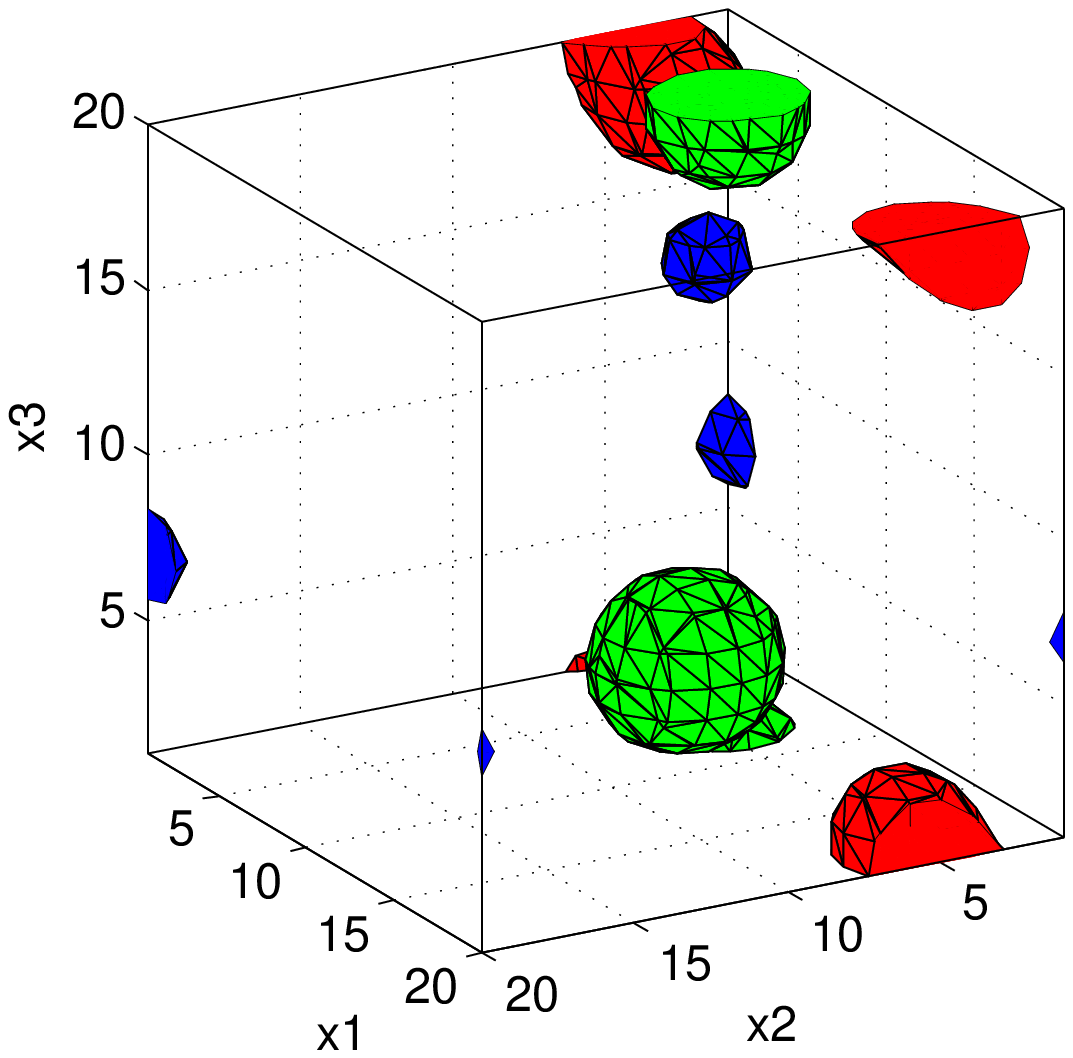}
\includegraphics[height=\fs]{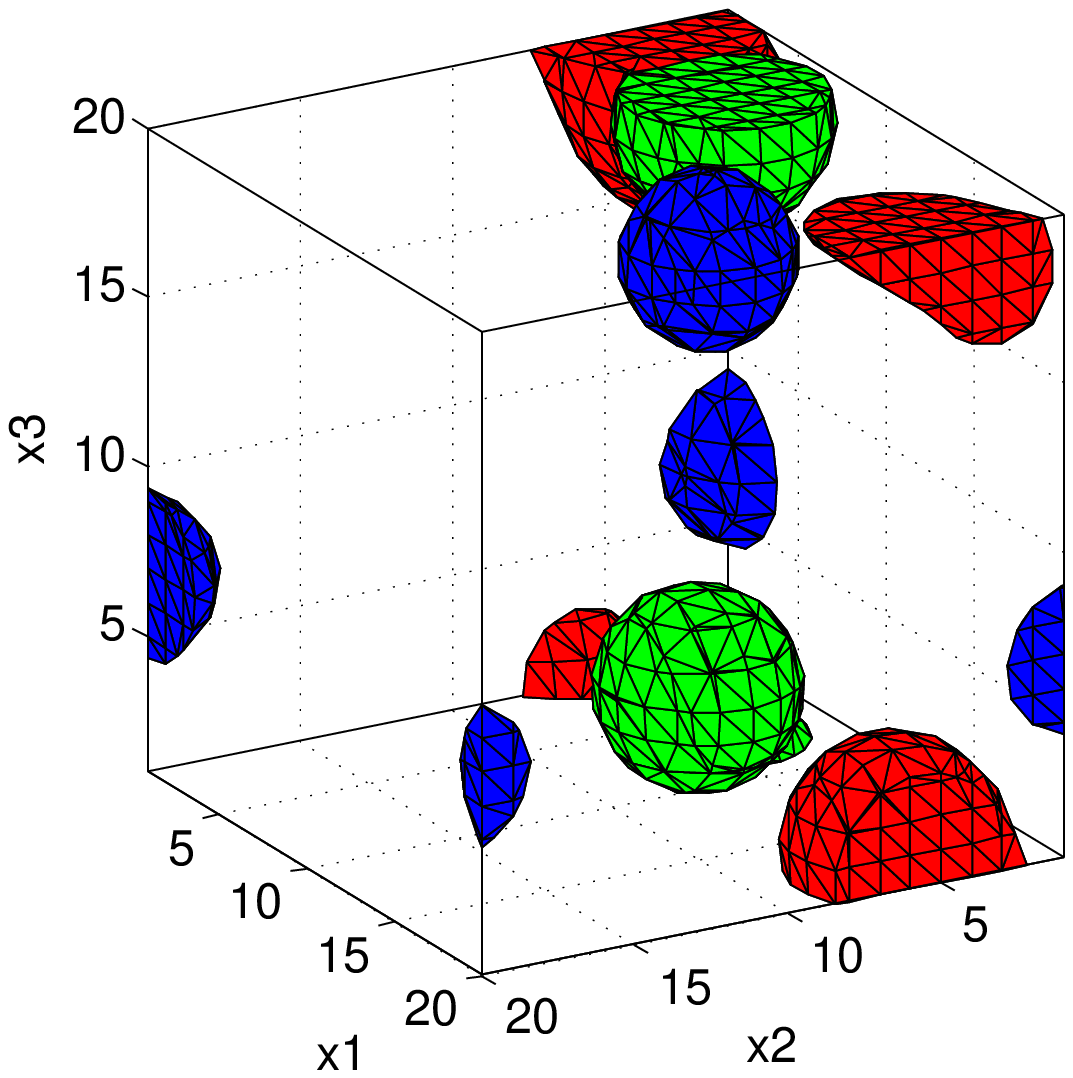}
\includegraphics[height=\fs]{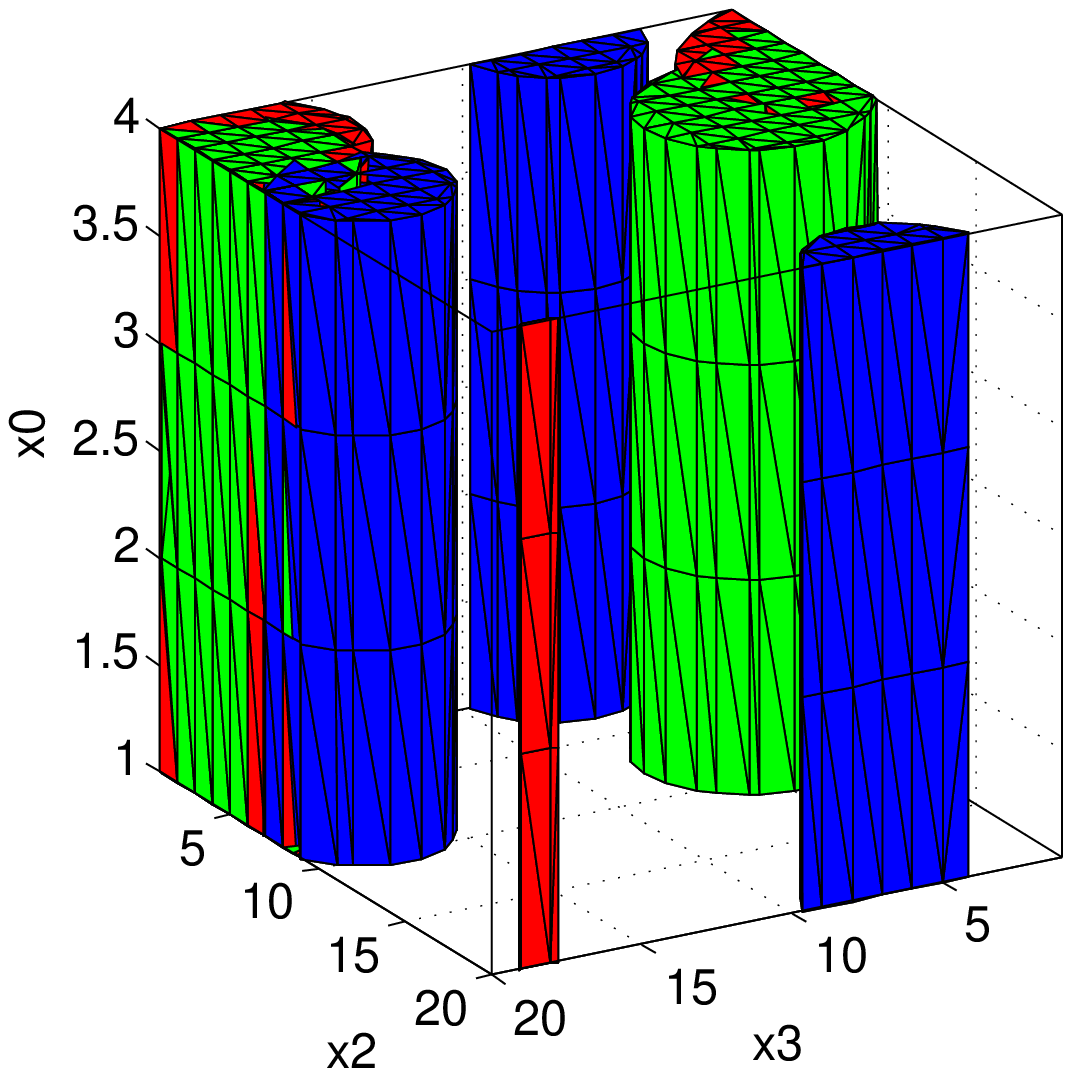}
\caption{Caloron (example 2) with non-trivial holonomy and $~|Q_\tx{t}|=2$;
  six constituents are visible in all plots: action density (u.l.), monopoles 
  by degenerate eigenvalues of $~\Pol(\vec{x})~$ (u.r.);
  $~\Psi_z^\dagger\Psi_z~$ for $~z=1/6, 1/2, 5/6~$ (different colors/brightness)
  (l.l. and l.r.).}   
\label{fig:q2}
\end{figure}

The second example of a selfdual lattice gauge field has $~Q_\tx{t}=2~$ and
also possesses a non-trivial holonomy (cf. \tref{tab:gluon2}). 
\begin{table}[tbp]
\centering
\begin{tabular}{|cc|}
\hline
\hline
\multicolumn{2}{|c|}{example 2 -- global observables} \\
\hline
\hline
$S/S_\tx{inst}$      & $\phantom{+}2.04$ \\
$Q_\tx{t}$ & $\phantom{+}2.00$ \\
$\mu_i$      & $\left\{\begin{array}{r}-0.35\\ -0.01\\ 0.36\end{array}
\right\}$\\
$n_+-n_-$    & $0-2$\\
$\delta_t$   & $7.4\,\cdot10^{-3}$\\
$\delta_F$   & $0.04$\\
\hline
\hline
\end{tabular}
\caption{Properties of a $~Q_\tx{t}=2~$ caloron gauge field
shown in Figs. \ref{fig:q2} and \ref{fig:q2flow}.}
\label{tab:gluon2}
\end{table}
It is obtained by cooling with respect to Wilson gauge action and by
applying stopping condition \ref{stop2}. Again the 
expected full number of  $~3~|Q_\tx{t}|=6~$ monopoles
can be identified by the {\it monopole criteria} {(b--d)} -- note that there
are only 5 separate massive lumps visible as separate maxima 
in the {\it isosurface plot of
the action density} and an application of criterion (a) is more
subjective~\footnote{A more sophisticated 
cluster algorithm for example, would be able to find 
six clusters of topological charge}. The decomposition into six monopole
constituents can be seen in \fref{fig:q2}.

\begin{figure}
\centering
\includegraphics[height=\fs]{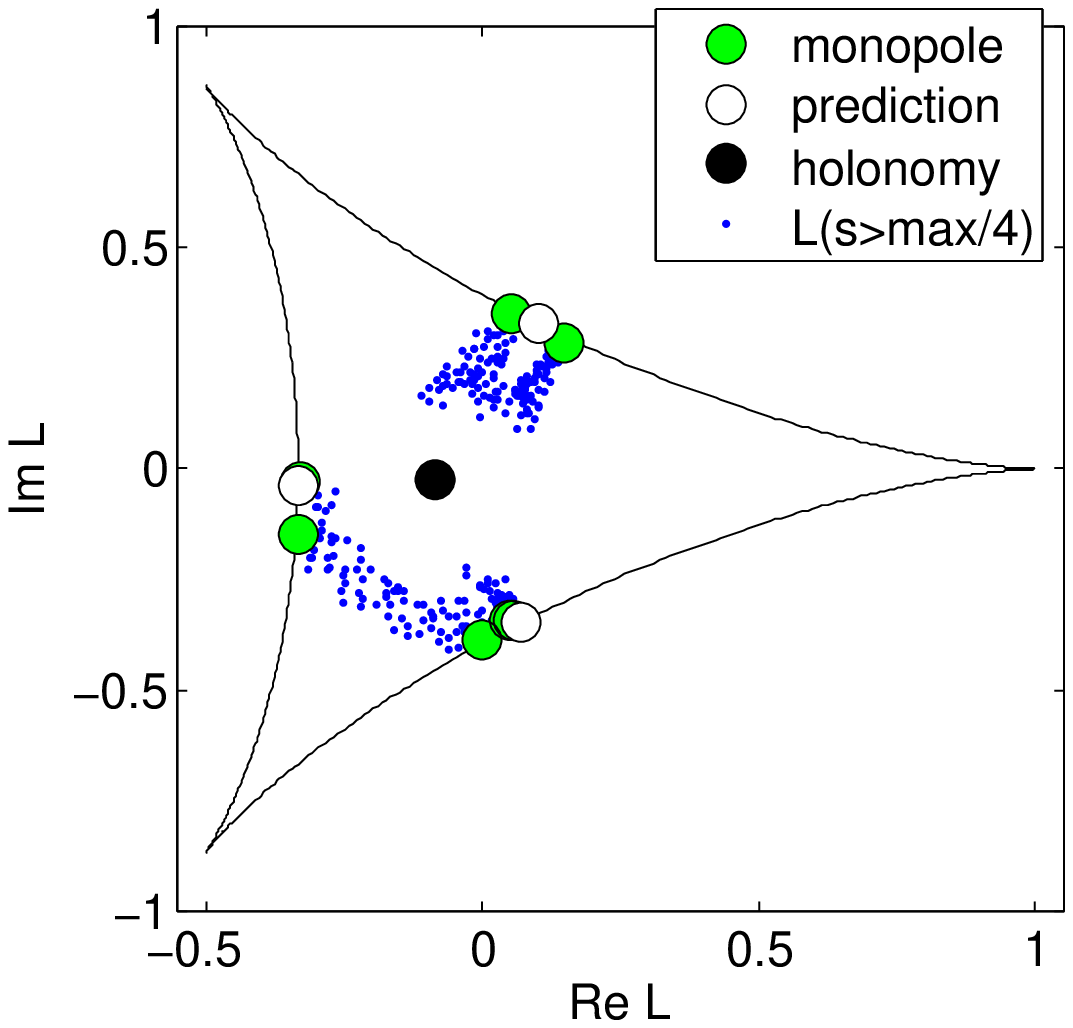}
\includegraphics[height=\fs]{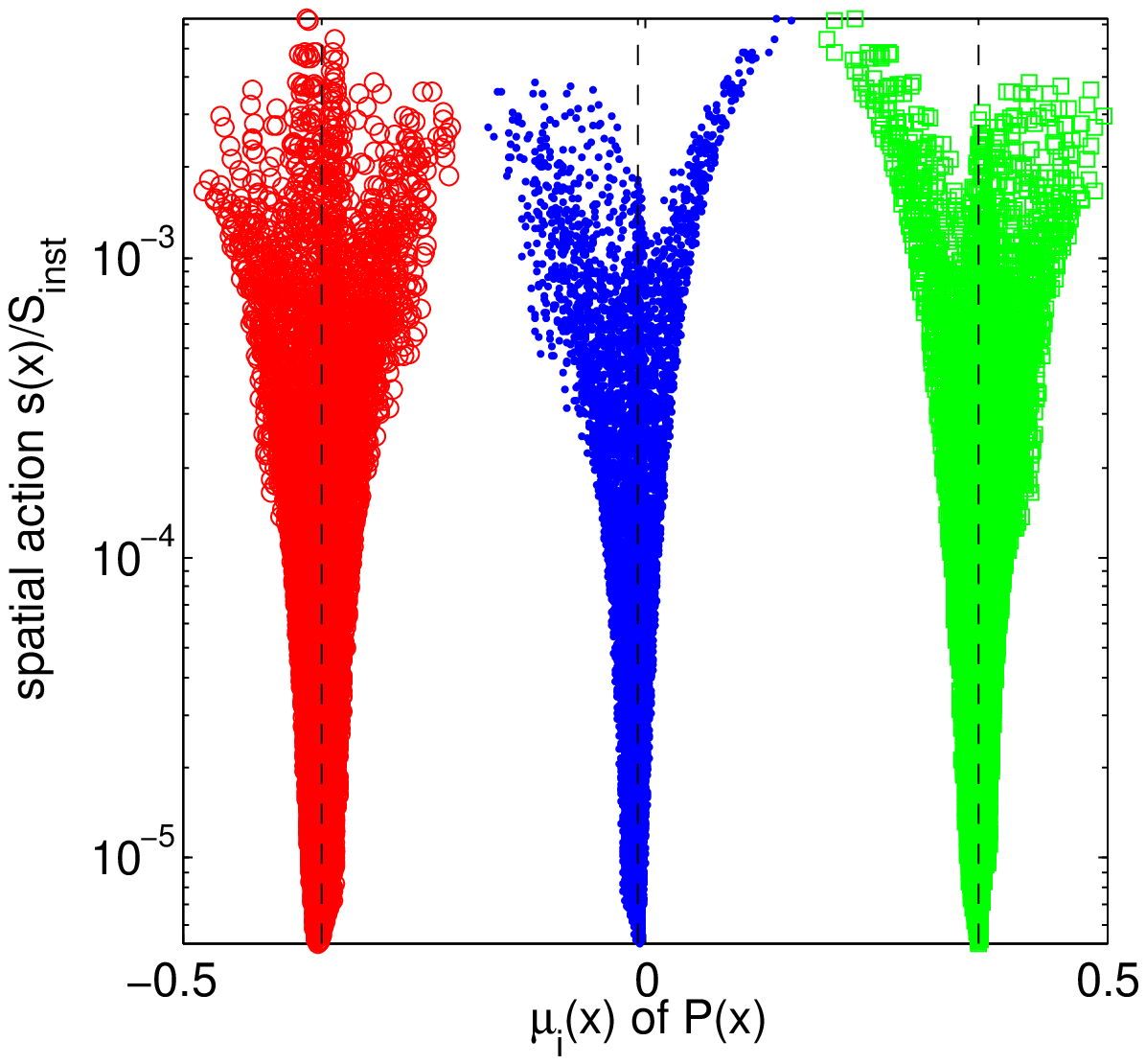}
\includegraphics[height=0.93\fs]{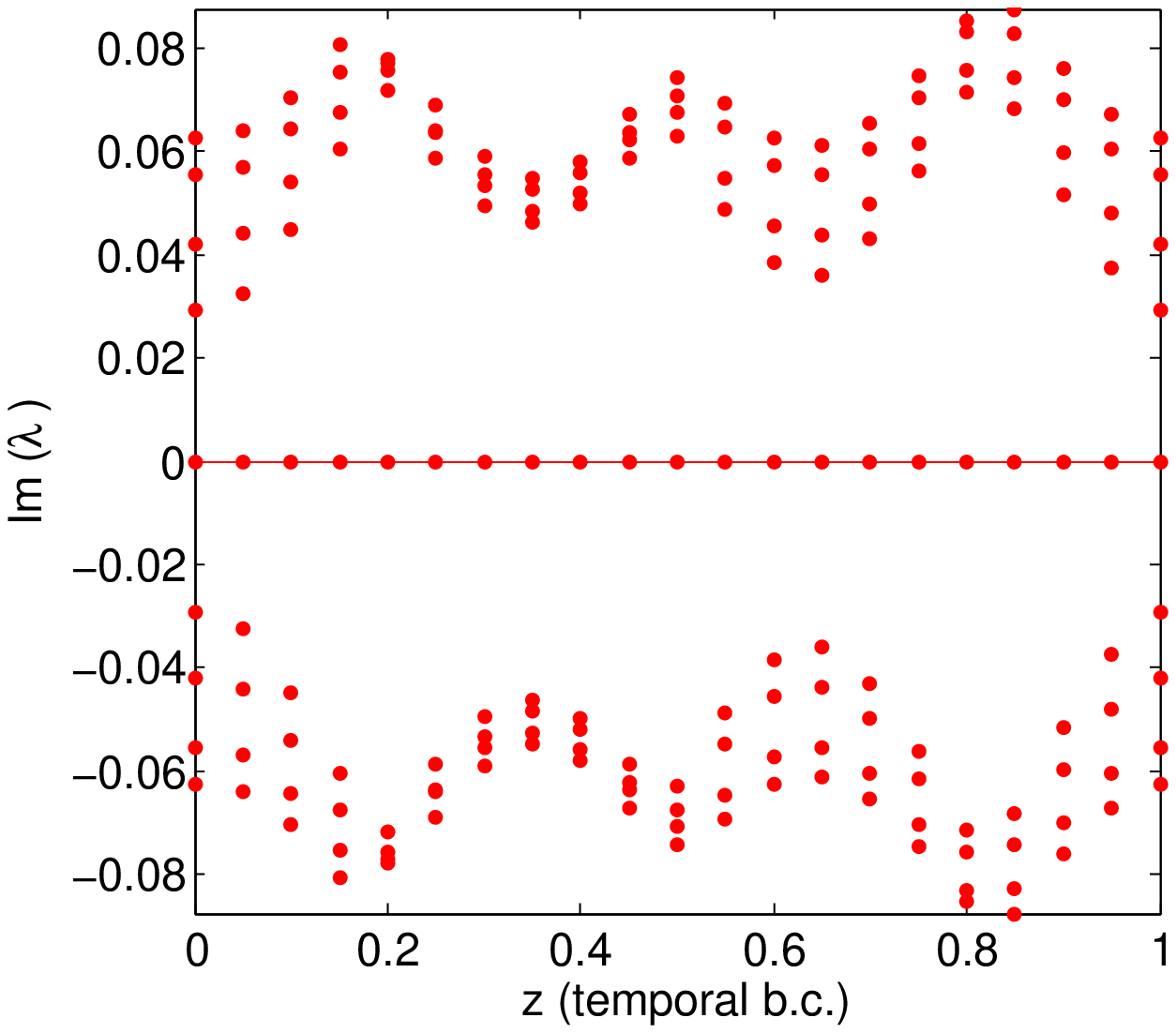}
\includegraphics[height=0.93\fs]{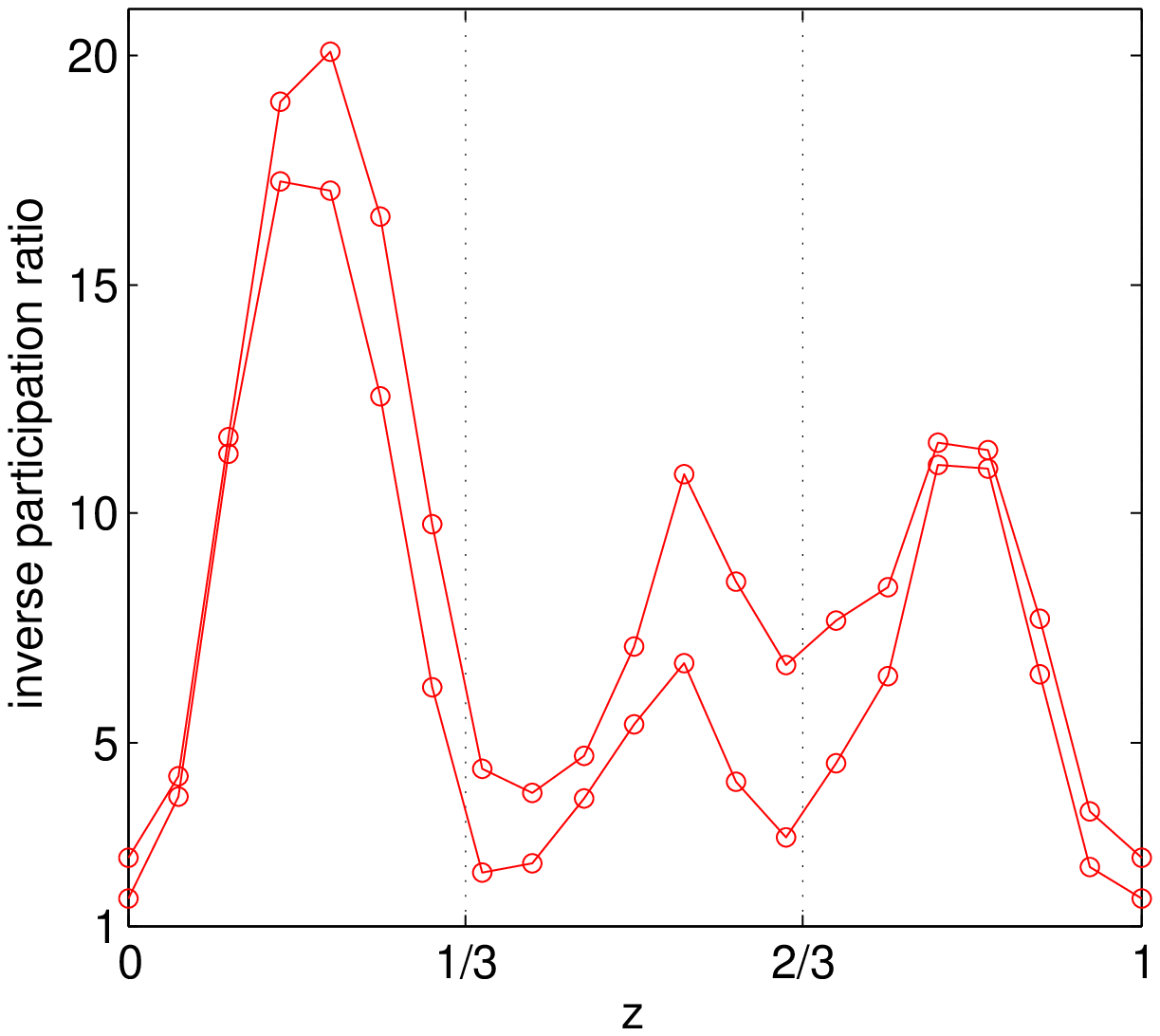}
\caption{Caloron (example 2); scatterplot of $\Pol(\vec{x})$ for 
  the points with high enough action density) in the
  complex plane (u.l.); eigenvalues of $\Pol(\vec{x})$ versus spatial
  action density (u.r.); $\Im(\lambda)$ (eigenvalue of $\slashed{D}$) (l.l.)
  and inverse participation ratio of the two zero-modes (l.r.) versus
  the temporal fermionic boundary condition $z$.}
\label{fig:q2flow}
\end{figure}

\noindent
By analyzing example 2 the results for $~|Q_\tx{t}|=1~$ are confirmed and 
generalized. Hence lattice calorons, especially those with non-trivial holonomy,
show the typical properties of KvB calorons. Further cooling reduces
the violation of the equation of motion and does not influence
the constituent positions too much. 
However, there is a striking difference to the case $~|Q_\tx{t}|=1~$ which
is a consequence of the Nahm duality \cite{Braam:1989qk},
which points out that classical fields with $~|Q_\tx{t}|=1~$ are unstable
on the torus. 
In agreement with the Atiyah-Singer index theorem the $~Q_\tx{t}=2~$ field
supports two zero-modes with negative chirality. The $z$-dependence 
of the inverse participation ratio $\IPR$ of $~\Psi_z^\dagger\Psi_z(x)~$
and the low lying eigenvalues of the Dirac operator are shown in the lower
part of \fref{fig:q2flow}. The behavior of both zero-modes -- 
the choice of basis is somewhat ambiguous --  is similar to the $~|Q_\tx{t}|=1~$
case. The Polyakov loop through the constituents aligns as it is
predicted by \mref{eqn:polinterplay}. The slight deviation, which can
be observed in the upper left part of \fref{fig:q2flow}, might be due
to finite volume effects. However, the upper right plot of the same
figure shows that the holonomy $\Hol$ is again well defined
(witnessed by the small variance of the eigenphases over the ``asymptotic'' 
subvolume $V_{\alpha}$).

\begin{figure}[t]
\centering
\includegraphics[height=\fs]{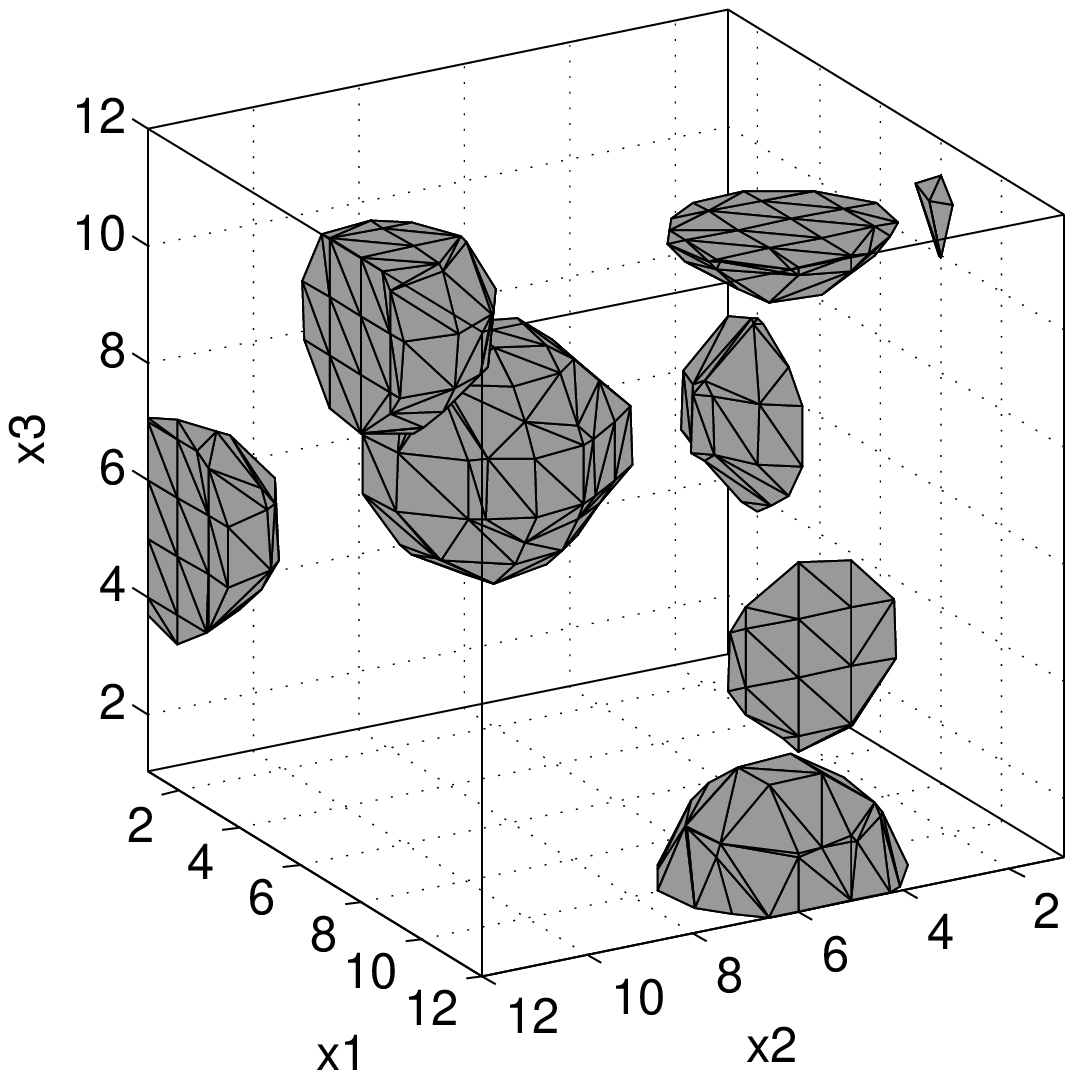}
\includegraphics[height=\fs]{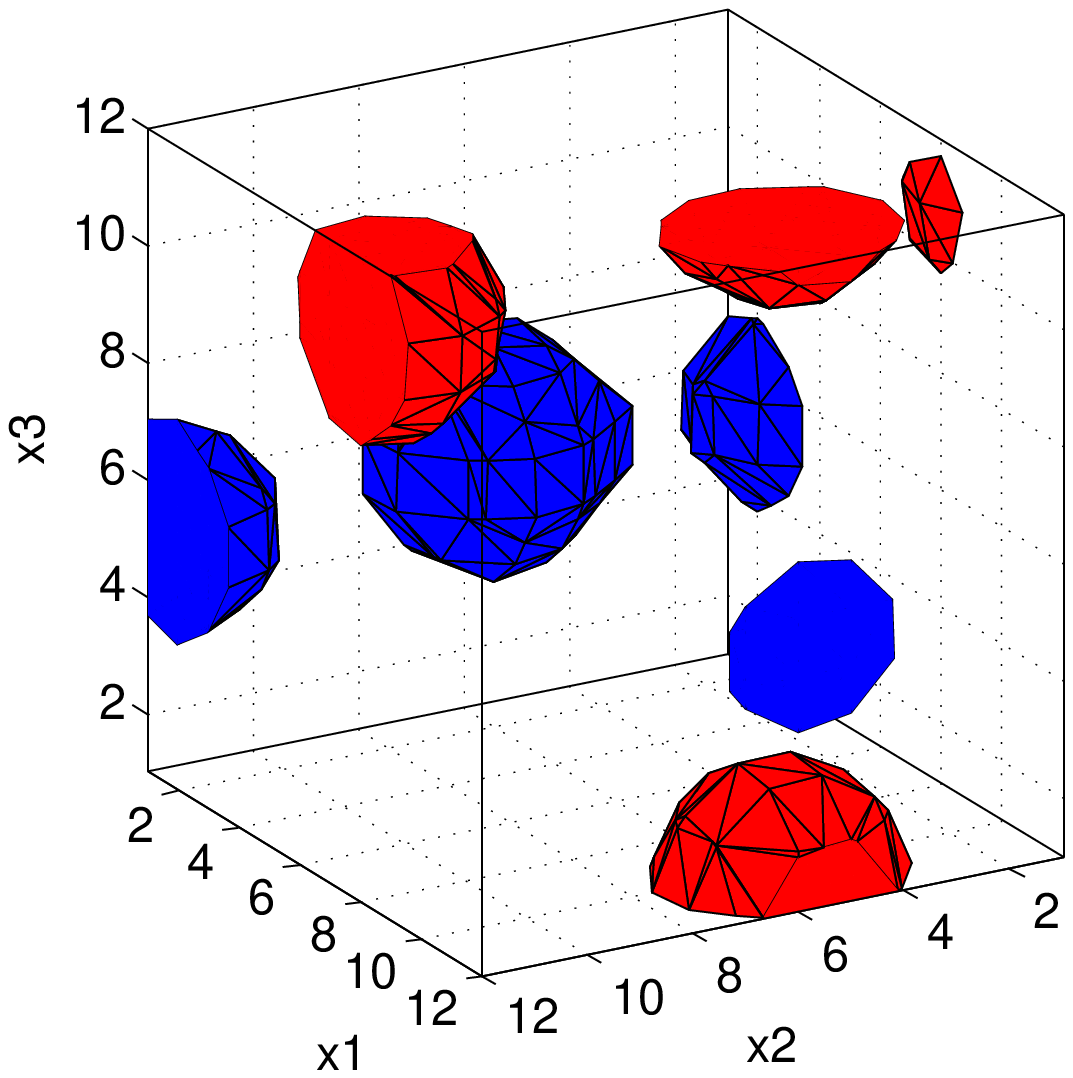}
\includegraphics[height=\fs]{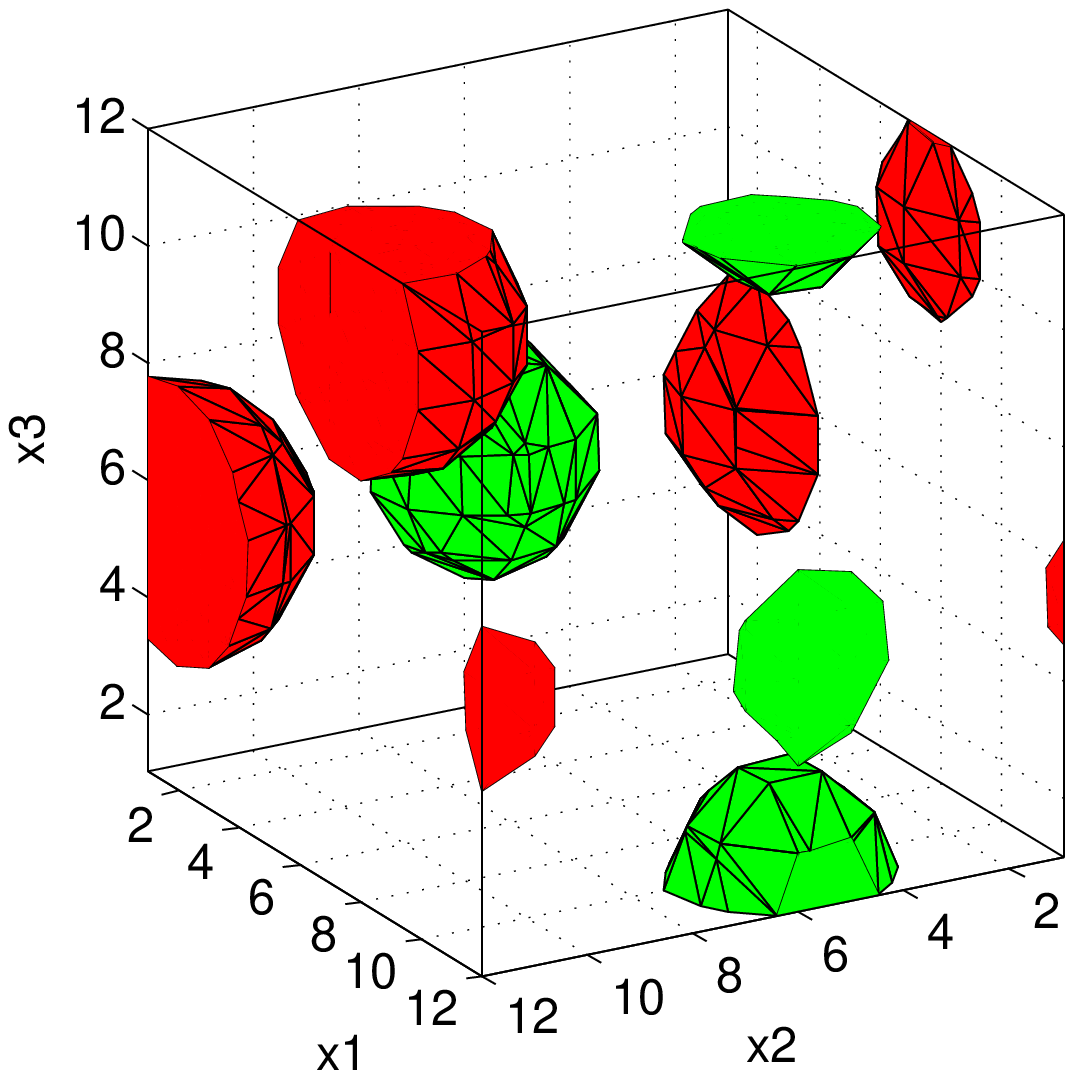}%
\includegraphics[height=\fs]{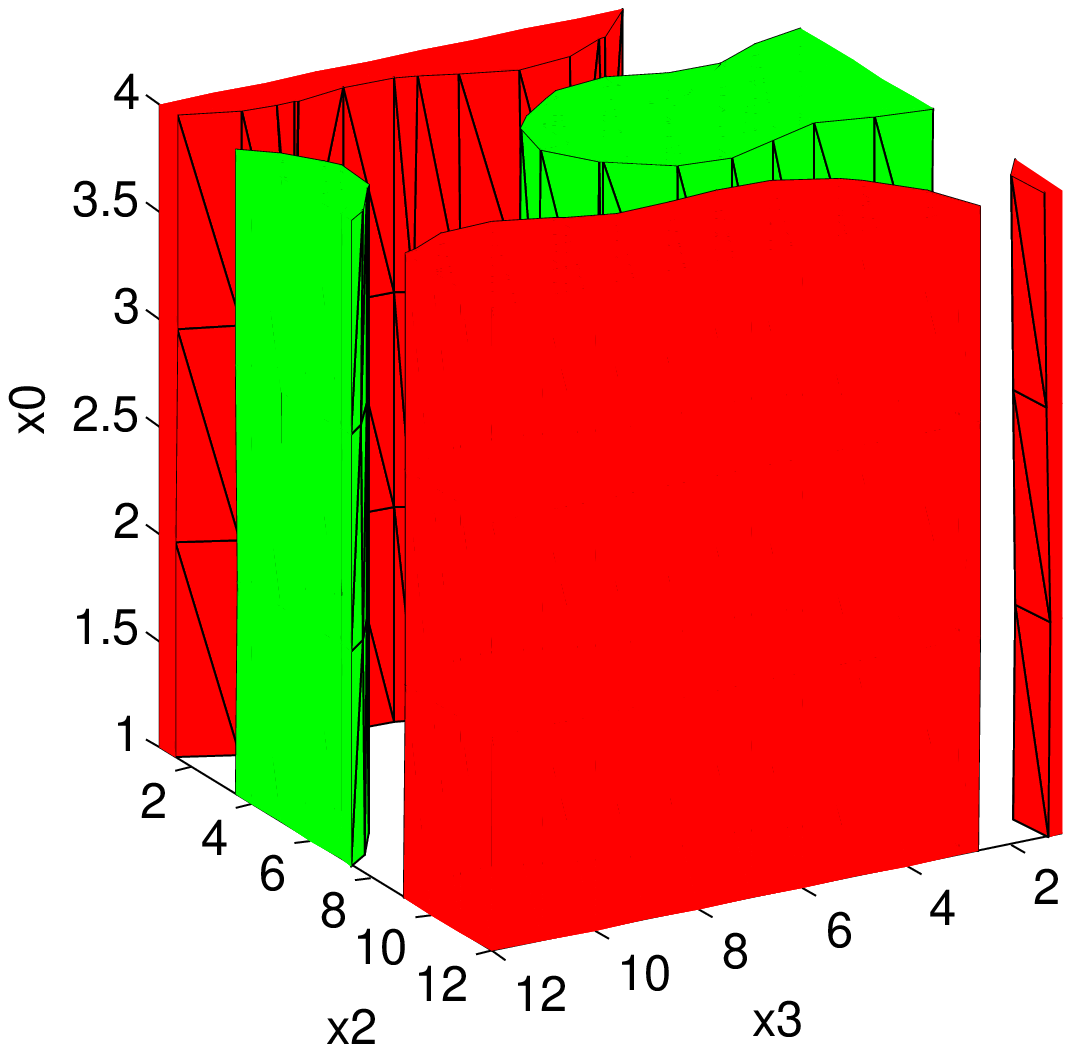}
\caption{Caloron-anticaloron superposition (example 3) with $~S=2 S_\tx{inst}~$,
  topological charge $~Q_\tx{t}=0~$; action density (u.l.), topological charge 
  density
  (red=positive, blue=negative) (u.r.); $~\Psi_z^\dagger\gamma_5\Psi_z(x)~$
  for $~z=0,0.5~$ (in different colors/brightness for each would-be zero-mode),
  shown in $3D$ space (l.l.) and depending on Euclidean time $x_0$ (l.r.).}
\label{fig:q0}
\end{figure}

The third and last example of an almost classical gauge field has 
$~Q_\tx{t}=0~$ but needs to be interpreted in terms of a caloron-anticaloron
superposition. The holonomy is not maximally non-trivial, giving rise
\begin{table}[t]
\centering
\begin{tabular}{|cc|}
\hline
\hline
\multicolumn{2}{|c|}{example 3 -- global observables} \\
\hline
\hline
$S/S_\tx{inst}$      & $\phantom{+}2.00$ \\
$Q_\tx{t}$ & $\phantom{+}0.00$ \\
$\mu_i$      & $\left\{\begin{array}{r}-0.18\\ -0.08\\ 0.26\end{array}\right\}$\\
$n_+-n_-$    & $0-0$\\
$\delta_t$   & $6.0\,\cdot10^{-2}$\\
$\delta_F$   & $0.18$\\
\hline
\hline
\end{tabular}
\caption{Properties of a caloron-anticaloron gauge field
shown in Figs. \ref{fig:q0} and \ref{fig:q0flow}.}
\label{tab:gluon3}
\end{table}
to a light, a moderately large and a heavy monopole constituent
(mass fractions $~\vec{\nu} \approx (0.1;0.35;0.55)~$ cf. \tref{tab:gluon3}). 
This $4\times 12^3$ gauge field has been cooled with respect to the standard
Wilson action and again obtained under the stopping condition \ref{stop2}. 
Only $~4 <3~|S/S_\tx{inst}|=6~$ constituents can be identified as massive 
lumps of action (a), which is not surprising keeping the monopole masses
$8\pi^2\vec{\nu}$ in mind. The upper panels of \fref{fig:q0} show the
isosurfaces of action density $~\bar{s}(x)~$ and the topological charge
density $~\bar{q}(x)$.   
The interpretation in terms of a caloron-anticaloron superposition is supported 
by the relatively large value of $~\delta_F=0.18~$ compared with example 2. 
Our choice of $~\delta_F < 0.2~$ barely allowed us to capture 
this approximate solution when the non-staticity $~\delta_t~$ was passing a 
minimum. Such fields can become ultimately stable ($~\delta_F\to 0~$) only
if the selfdual and anti-selfdual centers are infinitely separated. 
Any finite separation results in an attractive force between the
caloron and anticaloron that renders the superposition unstable.

\begin{figure}[t]
\centering
\includegraphics[height=\fs]{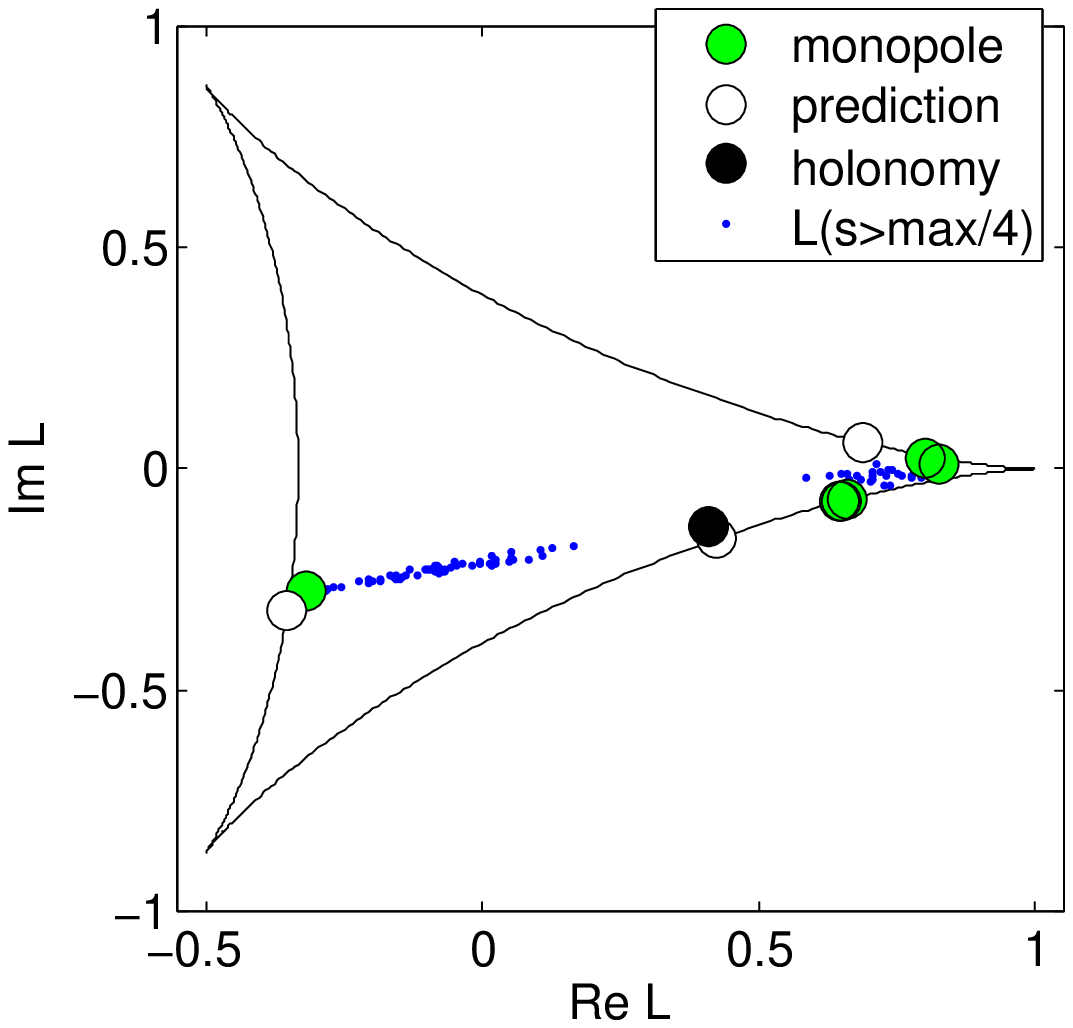}
\includegraphics[height=\fs]{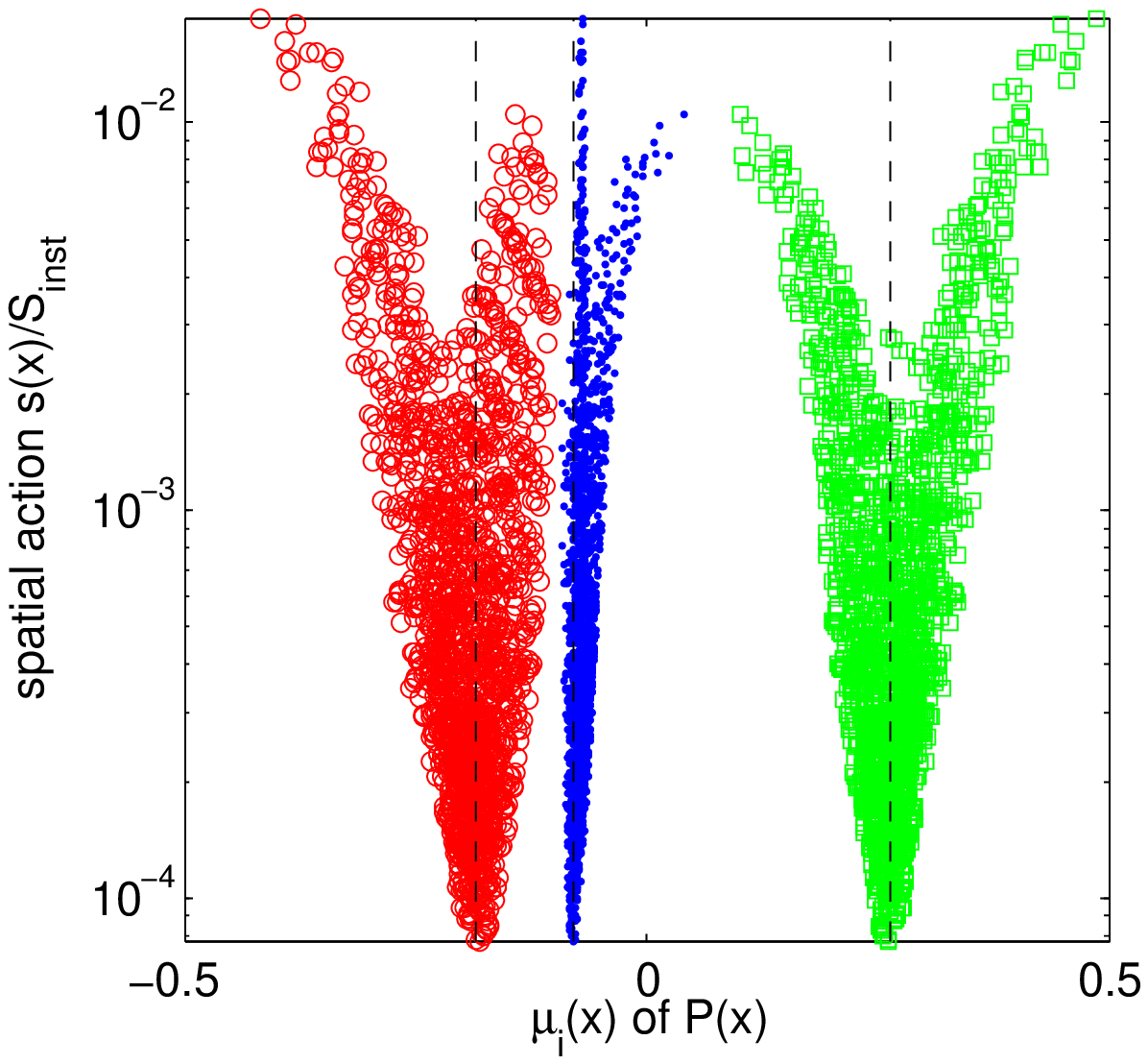}
\includegraphics[height=0.93\fs]{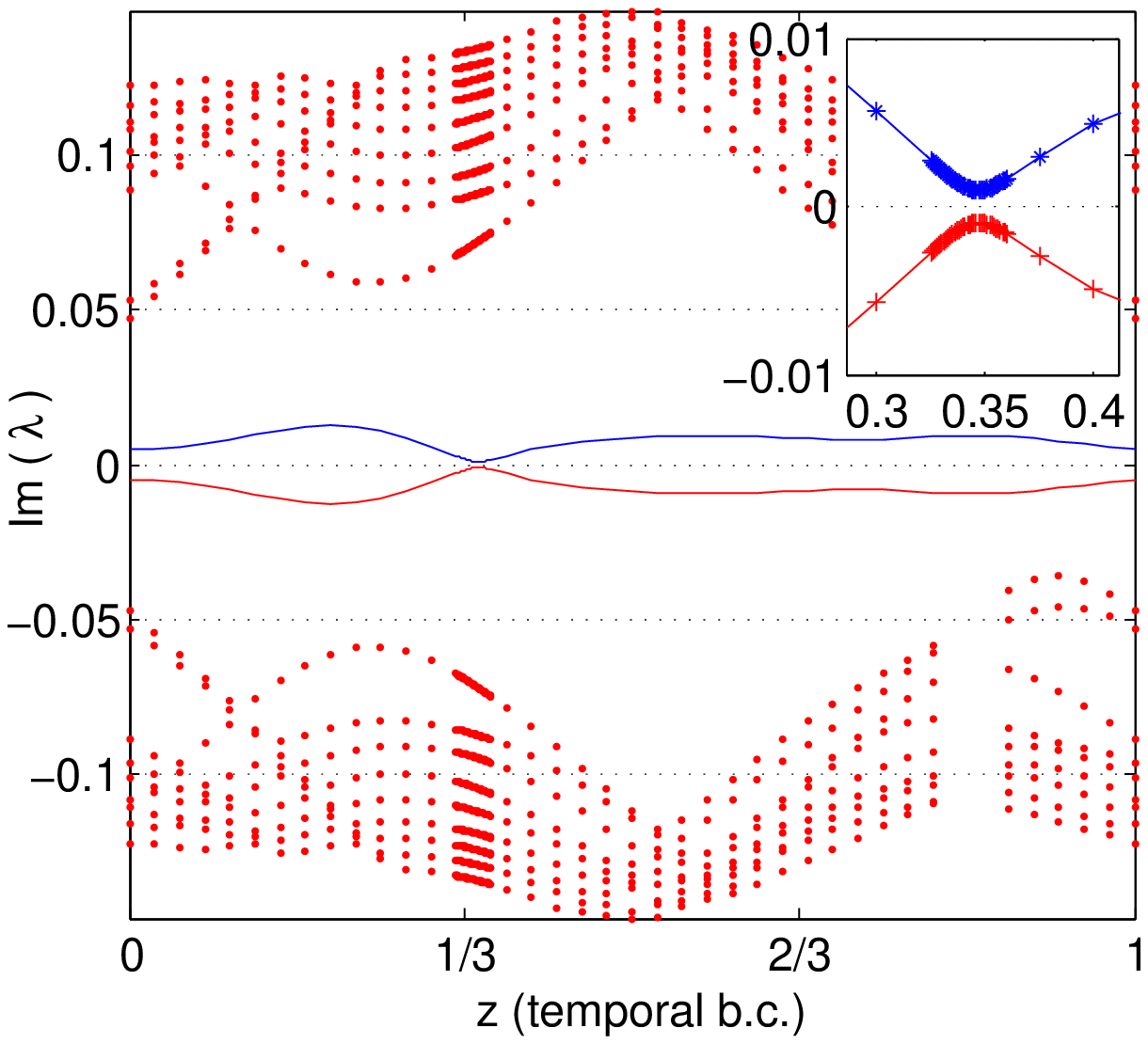}
\includegraphics[height=0.93\fs]{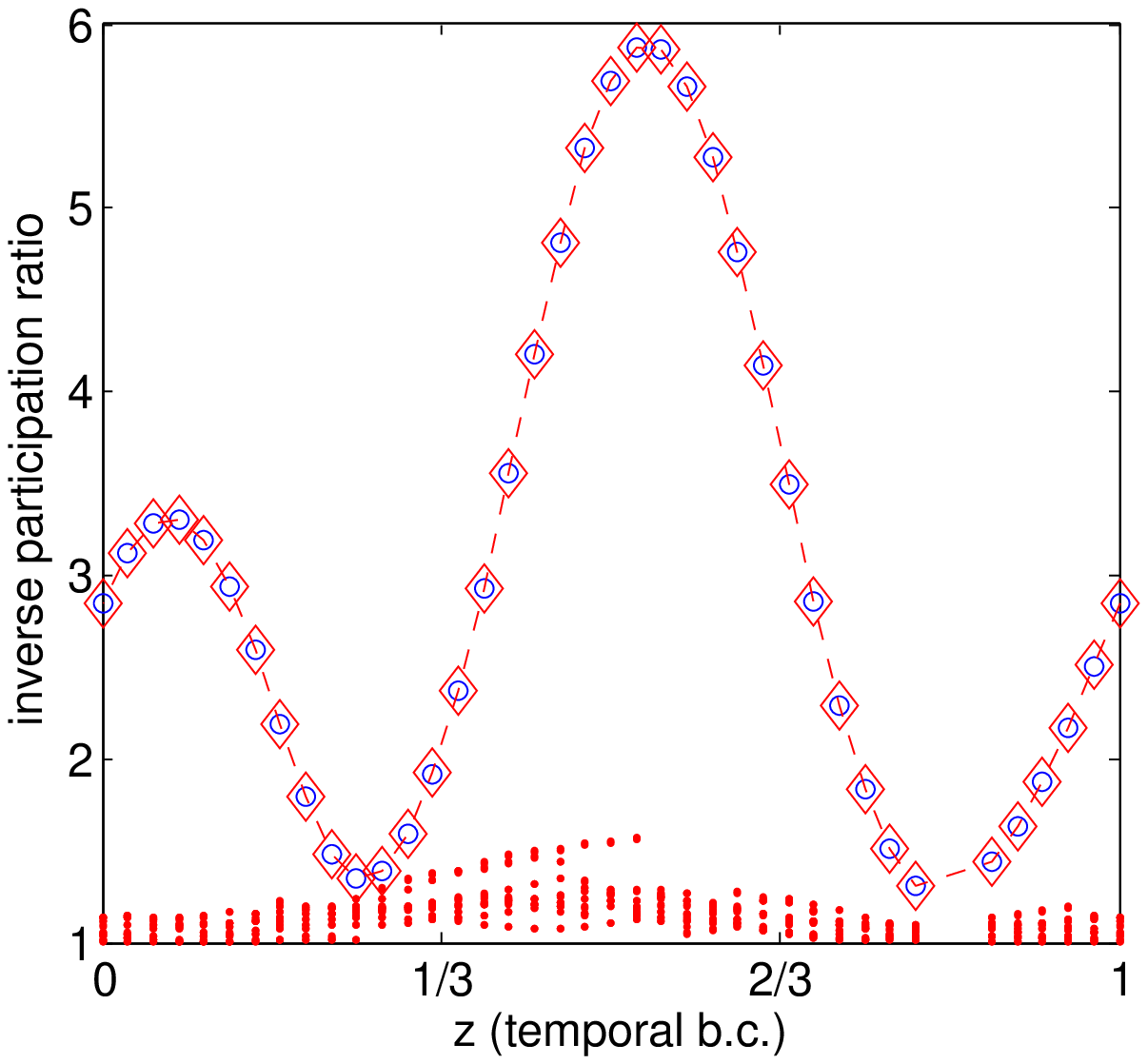}
\caption{Caloron-anticaloron superposition (example 3): scatterplot of
  $~\Pol(\vec{x})~$ (for the points with high enough action density) 
  in the complex plane (u.l.); eigenvalues of $~\Pol(\vec{x})~$ versus 
  spatial action density (u.r.); $~\Im(\lambda)~$ (eigenvalue of $~\slashed{D}~$) 
  (l.l.) and inverse participation ratio of the lowest eigenmodes (l.r.) 
  versus the temporal fermionic boundary condition $z$.} 
\label{fig:q0flow}
\end{figure}
There is no zero-mode present in this case. Instead we consider the
pair of non-zero modes (with the smallest non-vanishing imaginary part 
of the eigenvalue $\lambda$). For this pair of modes one needs to have
$~\langle\Psi_z|\gamma_5|\Psi_z\rangle=0~$. Hence, both eigenmodes, with
$~\Im\lambda~$ positive and negative, need to be localized simultaneously 
on the caloron and the anticaloron for a chosen boundary
condition. The characteristic jumping, presented for
$~|Q_\tx{t}|=1,2~$, can be seen for the positive or negative
part of $~\Psi_z^\dagger\gamma_5\Psi_z(x)~$. Both parts enclose
lumps of action with the same sign of topological density.
The flow of the inverse participation ratio $\IPR$ of the eigenmode density in
\fref{fig:q0flow} still shows two similarly localized eigenmodes
(cf. circles and diamonds in the lower right panel). In the same plot
it is visible that the modes only delocalize twice, which would
support an interpretation in terms of an embedded $SU(2)$ caloron--anticaloron. 
This is actually not the case, e.g. the scatterplot of the Polyakov
loop clearly exhibits a scattering in the full complex plane, which is
not possible for a trivial $SU(2)$ embedding. 
The interpretation in terms of a KvB caloron is still valid, however,
the behavior of a superposition of fields with opposite
duality\footnote{Only locally the field is \emph{mainly} either
selfdual \emph{or} antiselfdual.} differs from a fully selfdual or
anti-selfdual caloron. 

\section{Ensembles of lattice calorons}\label{calensembles}

\begin{figure*}
\centering
\includegraphics[height=1.4\fs]{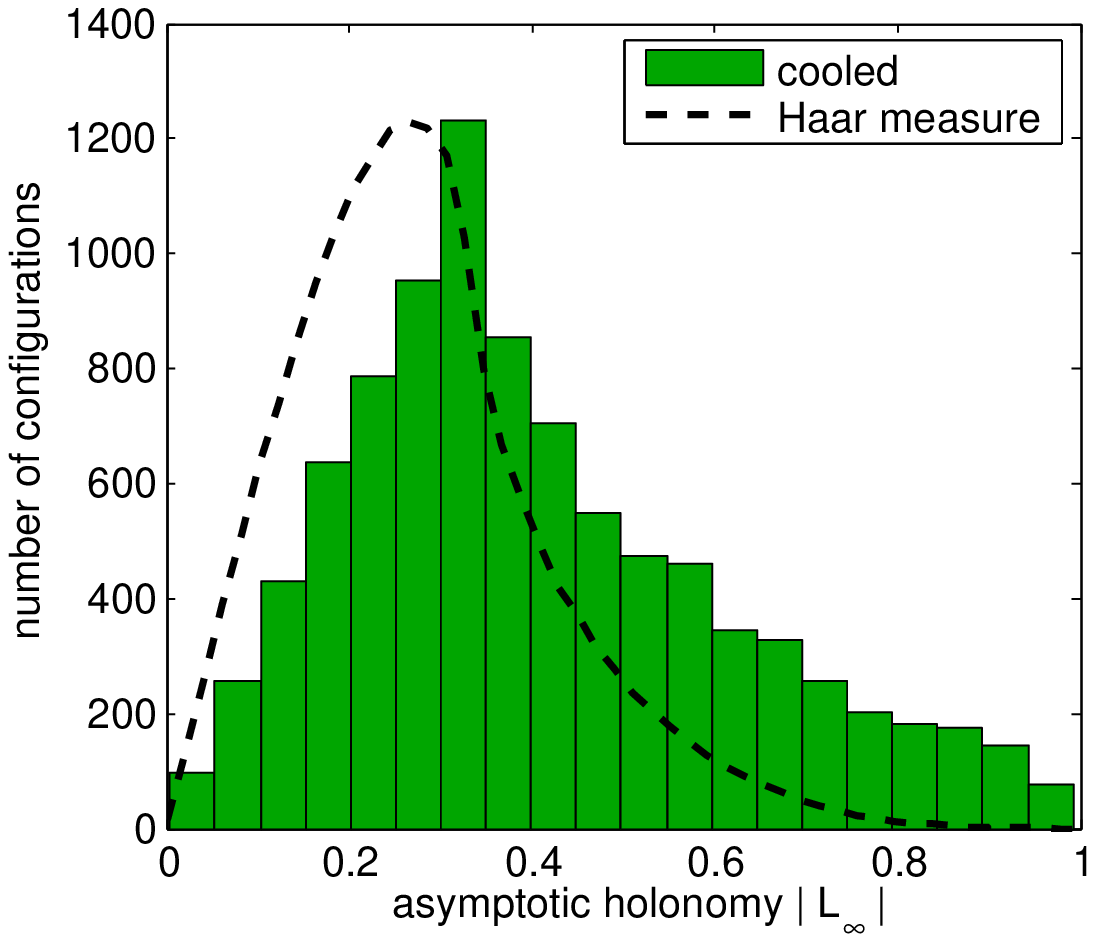}
\includegraphics[height=1.4\fs]{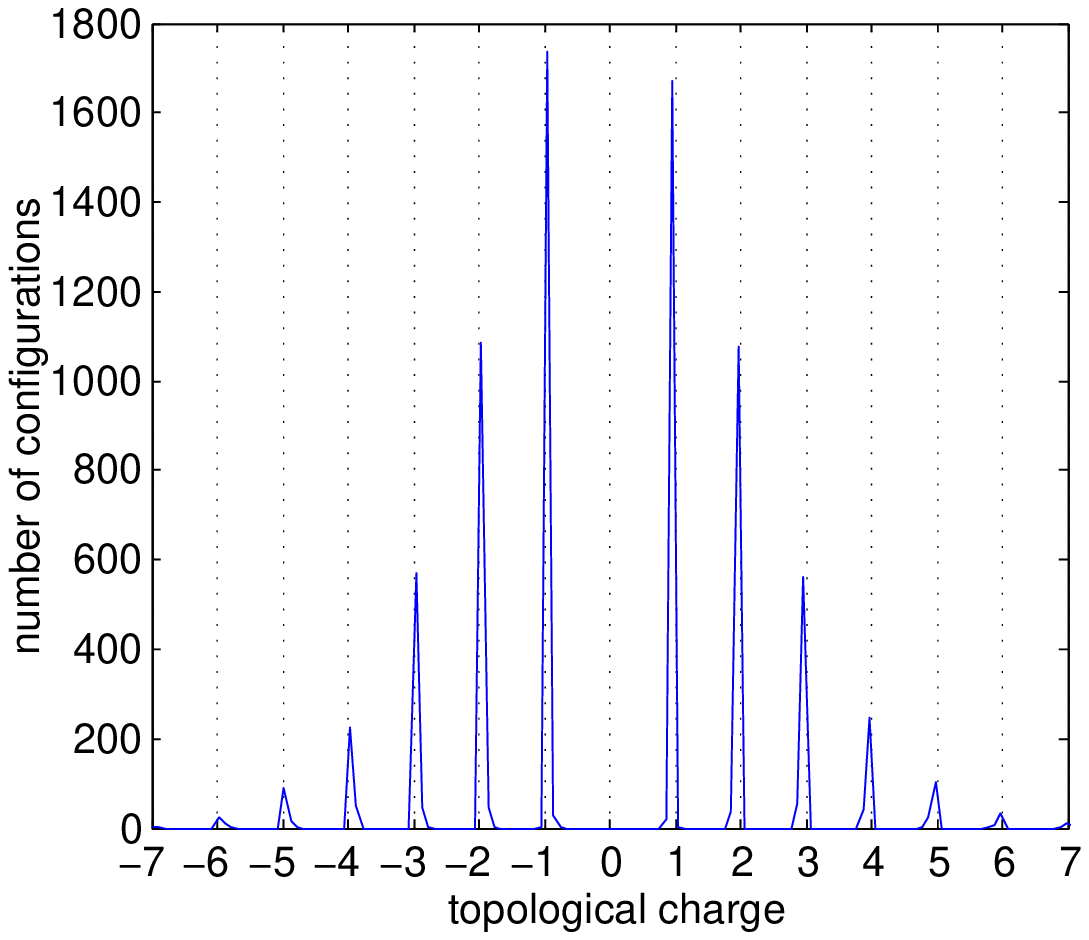}
\caption{Histogram of $|\tr\Hol|/3$ (left) and $Q_\tx{t}$ (right) for 
$4\times 12^3$ (stopping condition \ref{stop2}).}
\label{ens:distrib}
\end{figure*}

\begin{figure}
\centering
\includegraphics[height=0.93\fs]{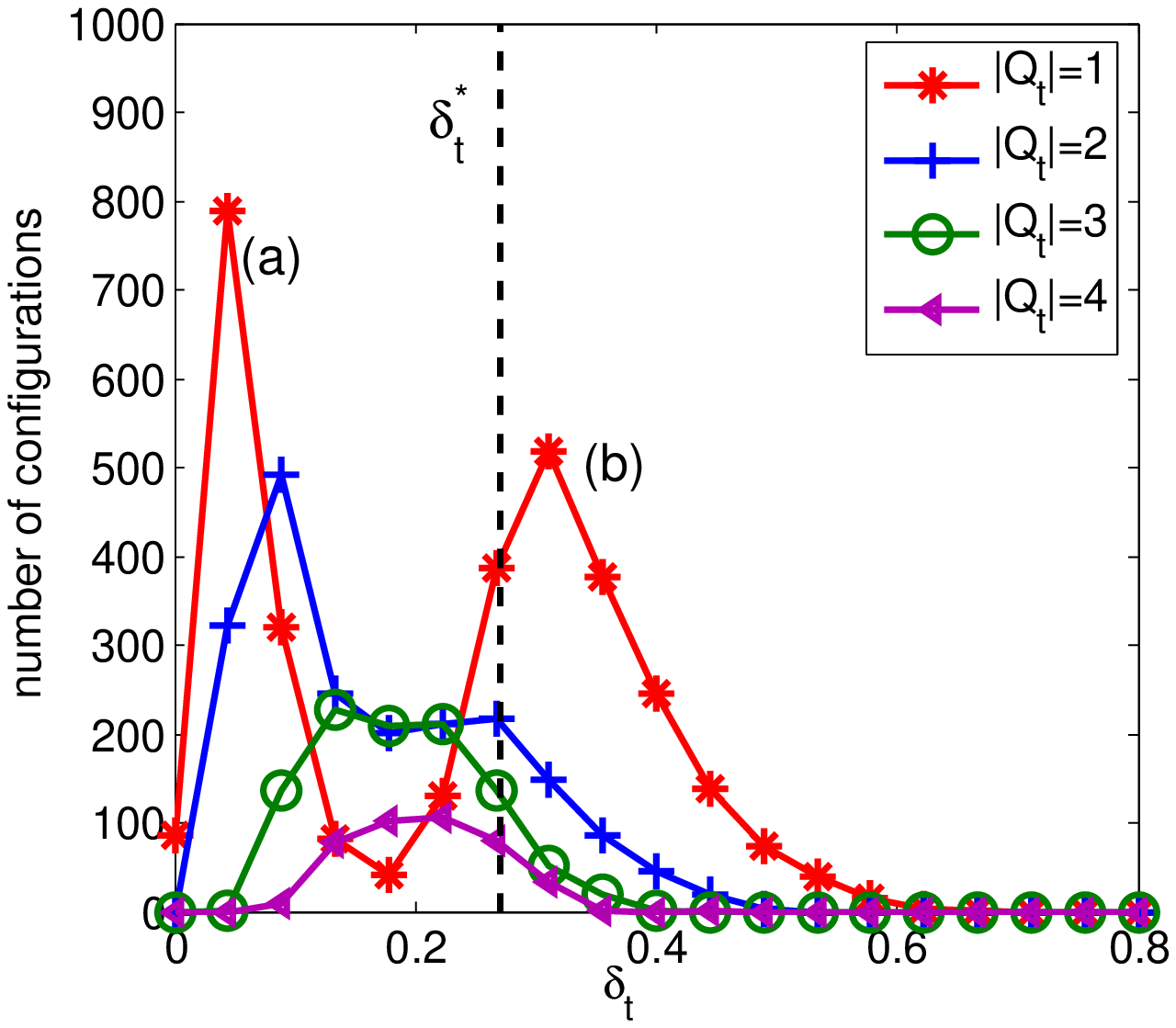}
\includegraphics[height=0.93\fs]{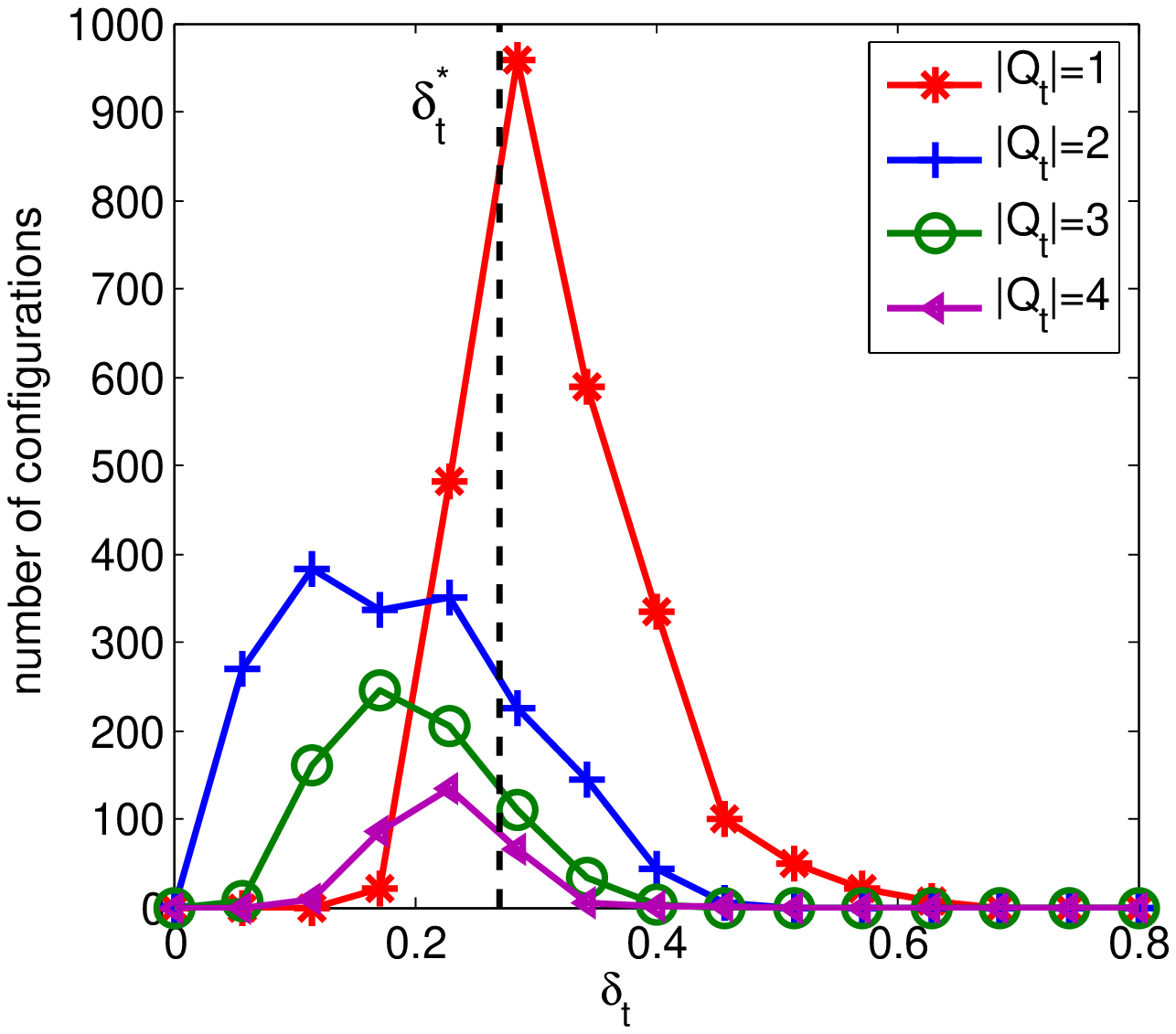}
\includegraphics[height=0.93\fs]{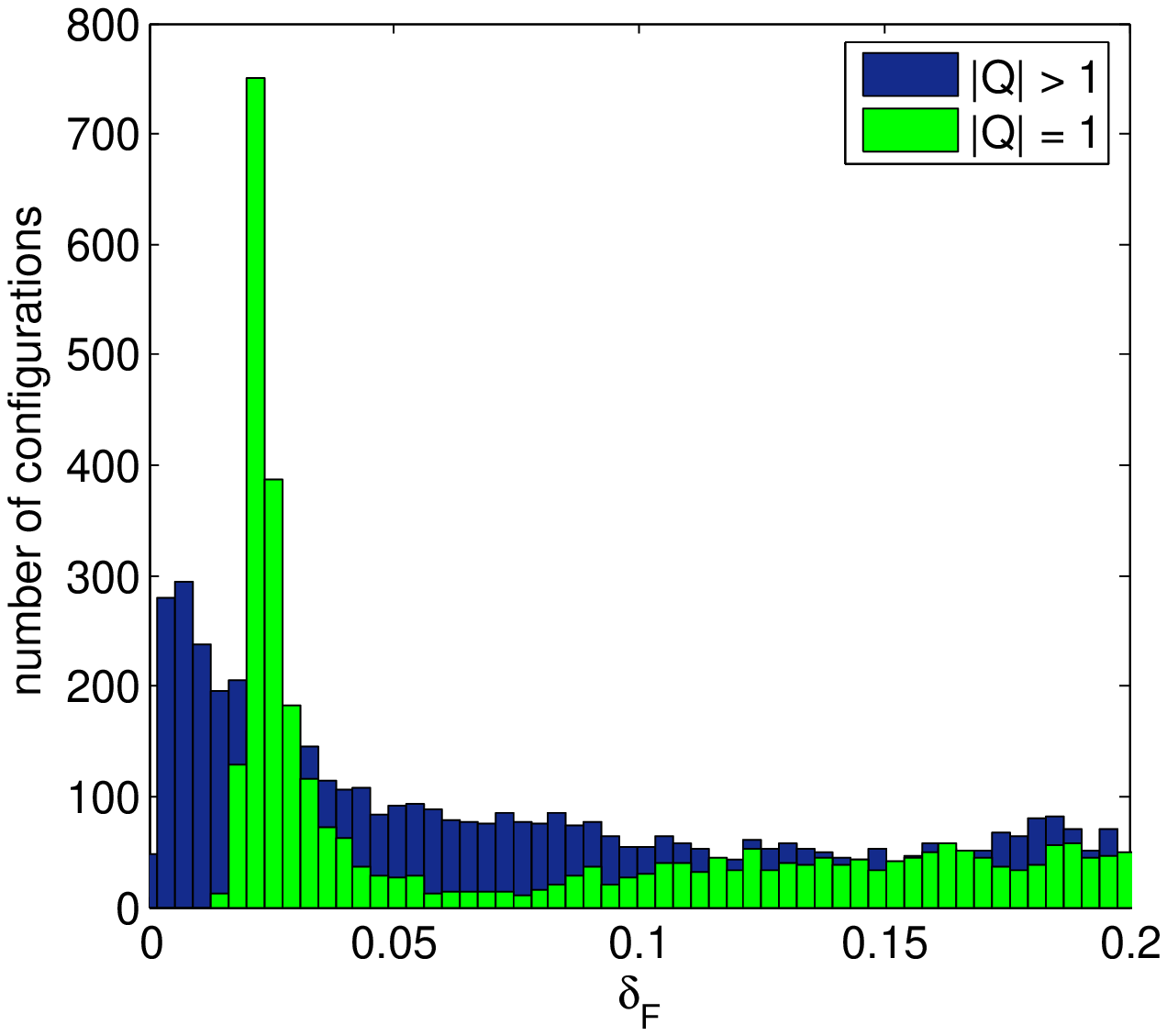}
\includegraphics[height=0.93\fs]{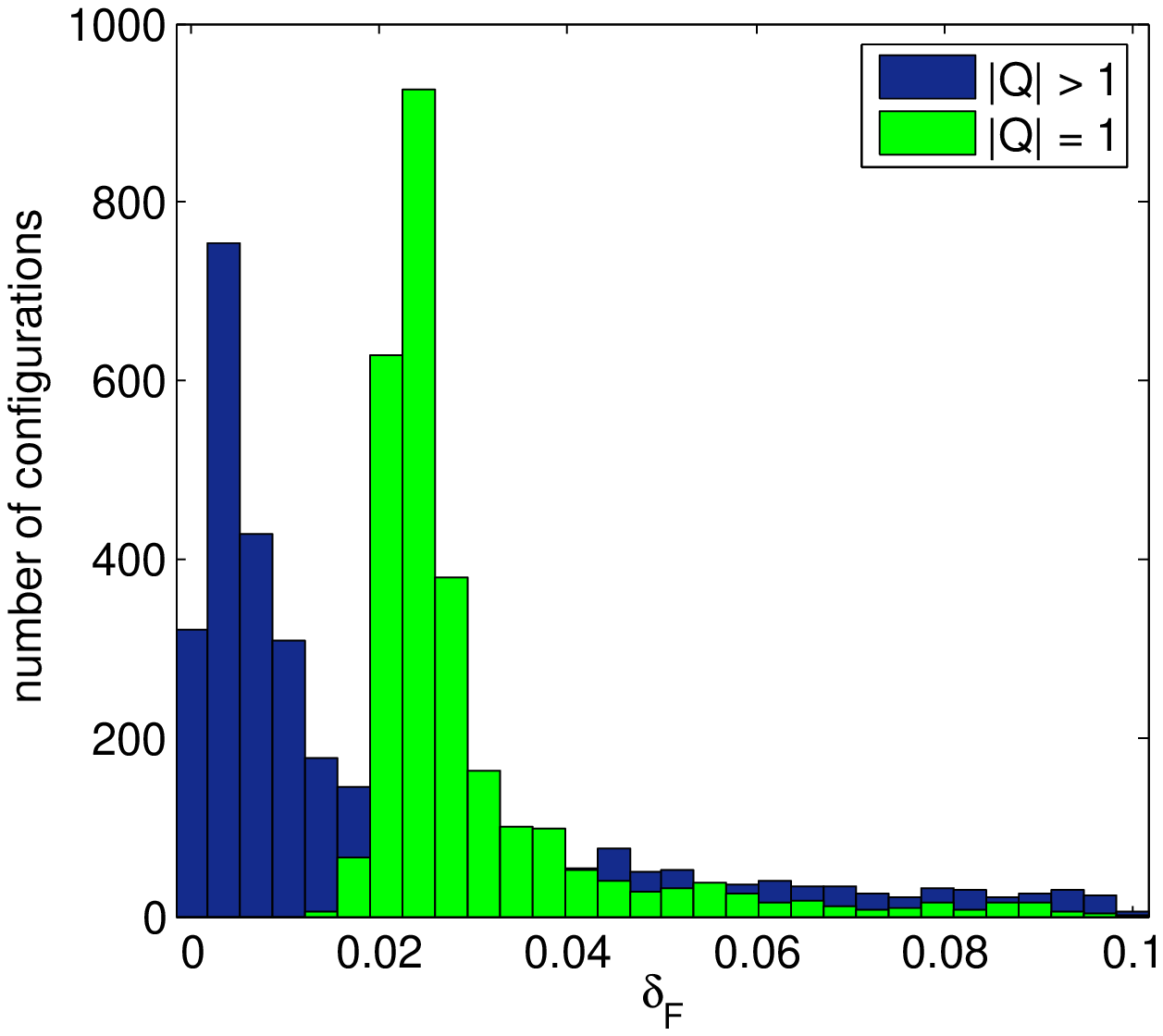}
\caption{Distribution of non-staticity $\delta_t$ (upper) and of the violation of
(anti) selfduality $\delta_F$ (lower) for stopping condition \ref{stop1} (right) 
and \ref{stop2} (left).}
\label{fig:distnonstat}
\end{figure}

To underline the importance of classical solutions with non-trivial
holonomy we show on the left in \fref{ens:distrib} a distribution of the 
modulus of the average of $~L(\vec{x})~$, which would be equal to one 
for trivial holonomy $\Hol\in Z(3)$ and smaller for non-trivial holonomy. 
It shows primarily that we are investigating
properties of classical fields with a non-trivial holonomy and that
these fields can be produced by cooling on the lattice. The
distribution of topological charges on the right in \fref{ens:distrib} 
shows the available range of topological sectors. 
The case $~Q_\tx{t}=0~$ is excluded,
because configurations of this class mostly pass cooling without being
stopped before they reach very low action (the third example above 
is a fortunate exception just tolerated by $~\delta_F < 0.2$). 
Both distributions only weakly depend on the stopping condition.
The distributions of the non-staticity $\delta_t$ and of the violation of 
selfduality ({\it i.e.} of the equation of motion) $\delta_F$
naturally depend on the stopping condition.
According to condition \ref{stop1} cooling proceeds until the equation of
motion is maximally fulfilled.  In this (actually late) stage the constituents 
are already squeezed. In the corresponding ensemble of cooled configurations 
one cannot find static calorons with separated constituents, at least for 
$~|Q_\tx{t}|=1~$ where squeezing is found to be a particularly strong effect. 
With the help of the non-staticity this role of the stopping criteria is 
visualized in the upper panels of \fref{fig:distnonstat} 
-- with stopping condition \ref{stop1} (upper right plot) no static calorons 
with $~|Q_\tx{t}|=1~$ ($\delta_t/\delta_t^*\ll 1$~) are found, 
whereas the combined stopping condition \ref{stop2} (upper
left plot) gives both static (peak (a)) {\it and} non-static calorons 
(peak (b)) with $~|Q_\tx{t}|=1$. 
This effect is much weaker for $~|Q_\tx{t}|=2,\ldots,4$. 

\noindent
For the general understanding it is also interesting to 
know how the violation of selfduality $\delta_F$ is distributed. For
both stopping criteria this is shown in the lower panels of
\fref{fig:distnonstat}. Clearly, the case $~|Q_\tx{t}|=1~$ appears
exceptional also in this respect since 
it has a positive lower bound for the violation of (anti)selfduality of 
$~\approx 0.02~$ in contrast to classical configurations with $~|Q_\tx{t}|>1$. 

\begin{figure*}
\centering
\includegraphics[height=1.4\fs]{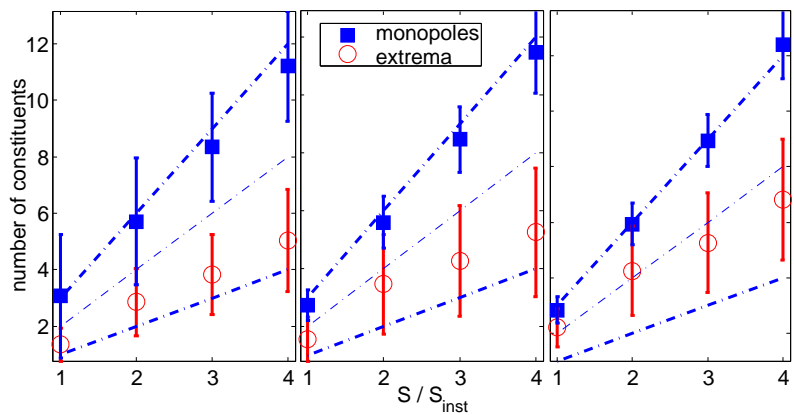}
\caption{Number of constituents for $4\times 12^3$ ensemble with
  standard deviation (bars); full squares by ``coinciding of
  eigenvalues'' criterion, open circles from local maxima in the 4D
  action density; (left) full ensemble; (middle) with
  $L_{\infty}<0.2$ and (right) with $\delta_t<0.1$ and
  $L_{\infty}<0.2$.}
\label{fig:nmono}
\end{figure*}

The previous figures describe the composition of {\it the ensembles}. 
In order to characterize better the  composition of {\it the configurations}
one should choose observables closer related to the KvB caloron
solution. It would be most natural to count the constituent monopoles
inside the caloron configurations. For this purpose we apply the
definition of a monopole as sitting at $\vec{Y}$ by two coinciding 
eigenvalues of $\Pol(\vec{Y})$, which was explained in
Chapter~\ref{latticiser}. Additionally we show  in \fref{fig:nmono}
the number of dominant 
({\it i.e.} $~|q_\tx{t}(Y)| > 0.1 \max_x |q_\tx{t}(x)|~$) 
local extrema in the topological charge density for the ensemble produced 
with stopping condition \ref{stop2}.

\noindent
Both methods, the number of monopoles provisionally defined as local maxima of the 
action/topological charge density or by our 
actually adopted monopole definition, expose (cf. in \fref{fig:nmono}) 
always more than $|Q_\tx{t}|$ lumps, a situation that would be expected for 
a caloron with massless or close-by constituents. With the actual monopole definition,
on average $3|Q_\tx{t}|$ monopoles are found in each topological sector. This
is independent of the cut, which can be applied to the ensemble to
make the monopole counting better defined\footnote{Without this cut
one finds a large number of spurious monopoles, in particular if the holonomy 
is not sufficiently non-trivial.}. By applying more strict cuts to the ensemble
(with respect to the holonomy and non-staticity, from left to right in 
\fref{fig:nmono}) the standard deviation of the number of monopoles for 
given $~S/S_\tx{inst}~$ decreases while on average the expectation to find 
$~3~|Q_\tx{t}|~$ monopoles is reconfirmed. 

\begin{figure*}
\centering
\includegraphics[height=1.4\fs]{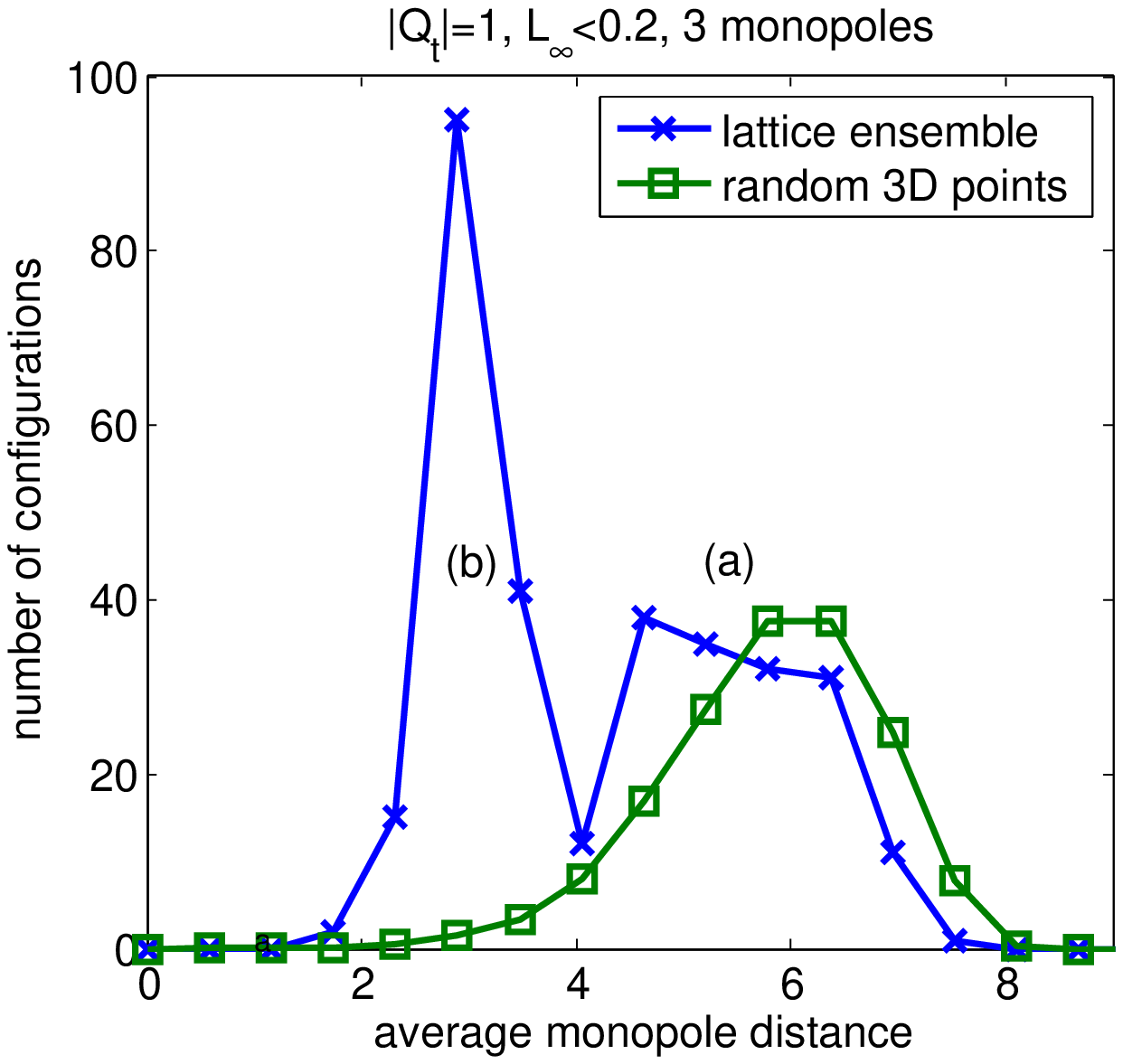}
\includegraphics[height=1.4\fs]{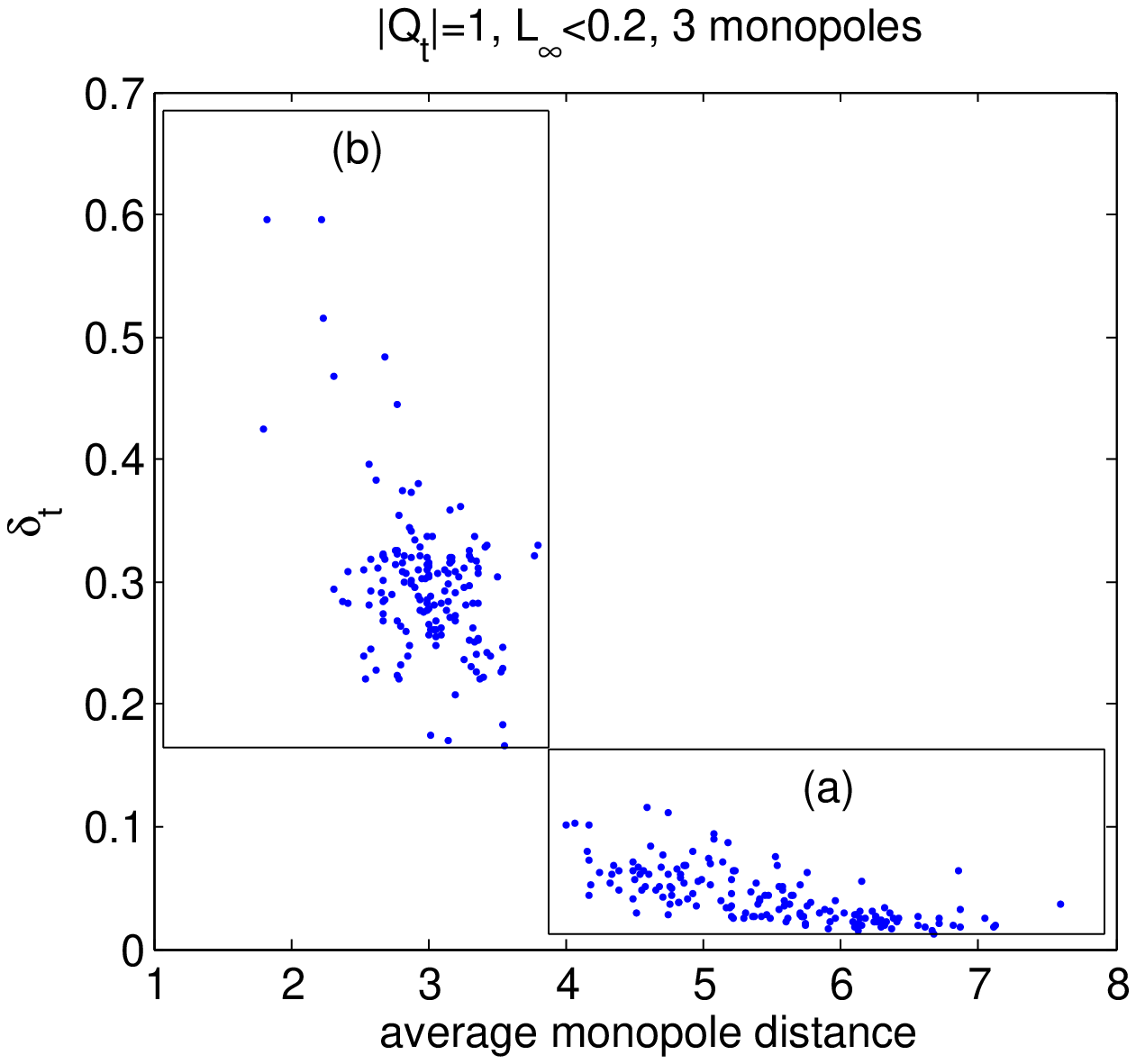}
\caption{Distribution of the average pairwise monopole distances
  compared with the result of an uniform distribution of 3D points
  (left). Average monopole distance versus the non-staticity
  $\delta_t$.} 
\label{fig:avgdist}
\end{figure*}

The splitting into static and non-static configurations in the
distribution of the non-staticity (cf. upper left in
\fref{fig:distnonstat}) for stopping condition \ref{stop2} is also
reproduced looking at the distribution of the average monopole
distance in \fref{fig:avgdist}. 
In this Fig. the sample is restricted to configurations with $|Q_t|=1$
where 3 monopoles have been observed.
The correspondence between the
clustering into static and non-static sets of configurations in
\fref{fig:distnonstat} and \fref{fig:avgdist} is highlighted by 
labelling those clusters of events (in both figures) as (a) and (b). 
They have emerged from the cooling process either because of 
the non-staticity went through a minimum (cluster (a)) 
or the violation of the equation of motion was minimal (cluster (b)).

\begin{figure}
\centering
\includegraphics[height=0.92\fs]{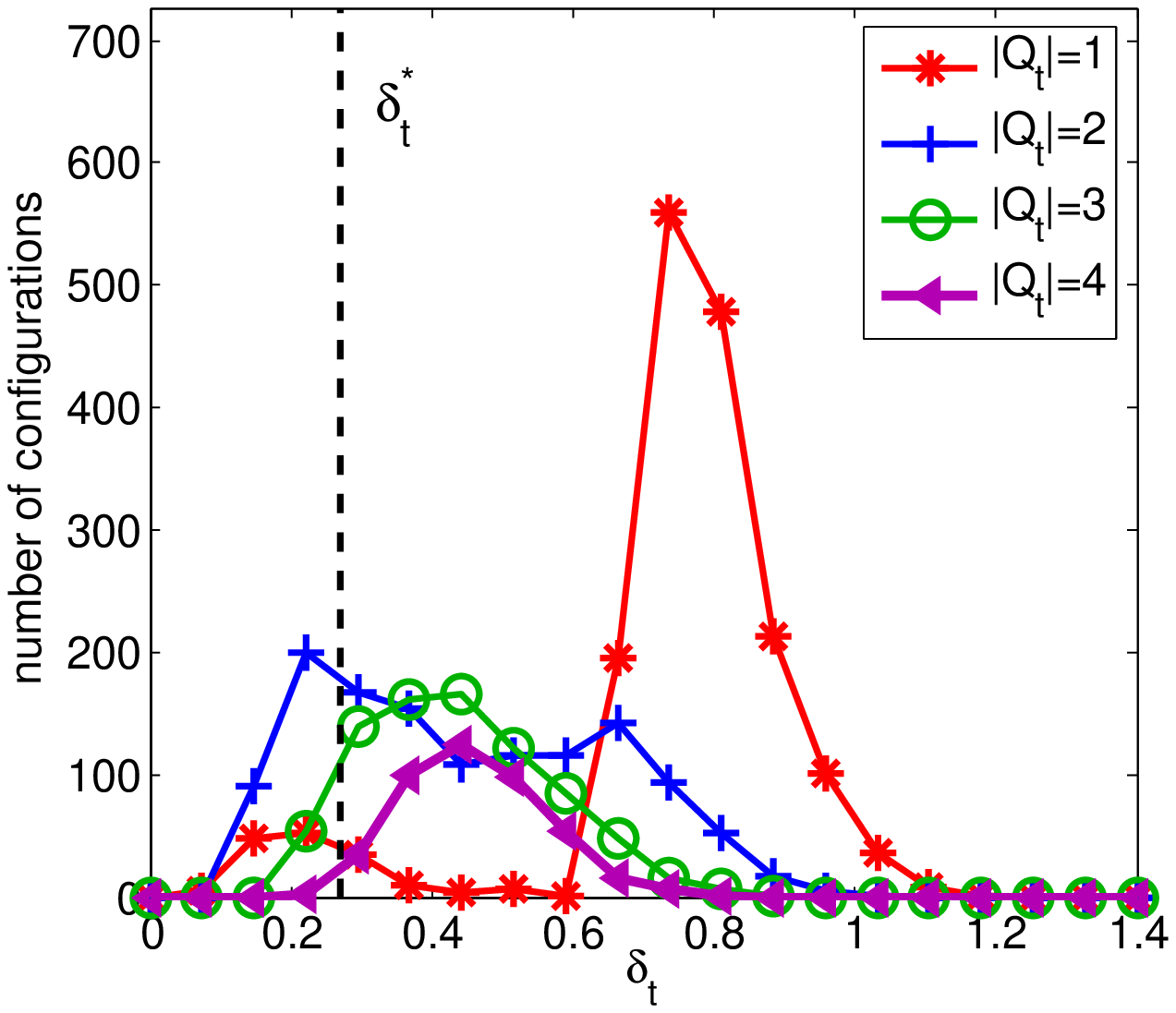}
\includegraphics[height=0.92\fs]{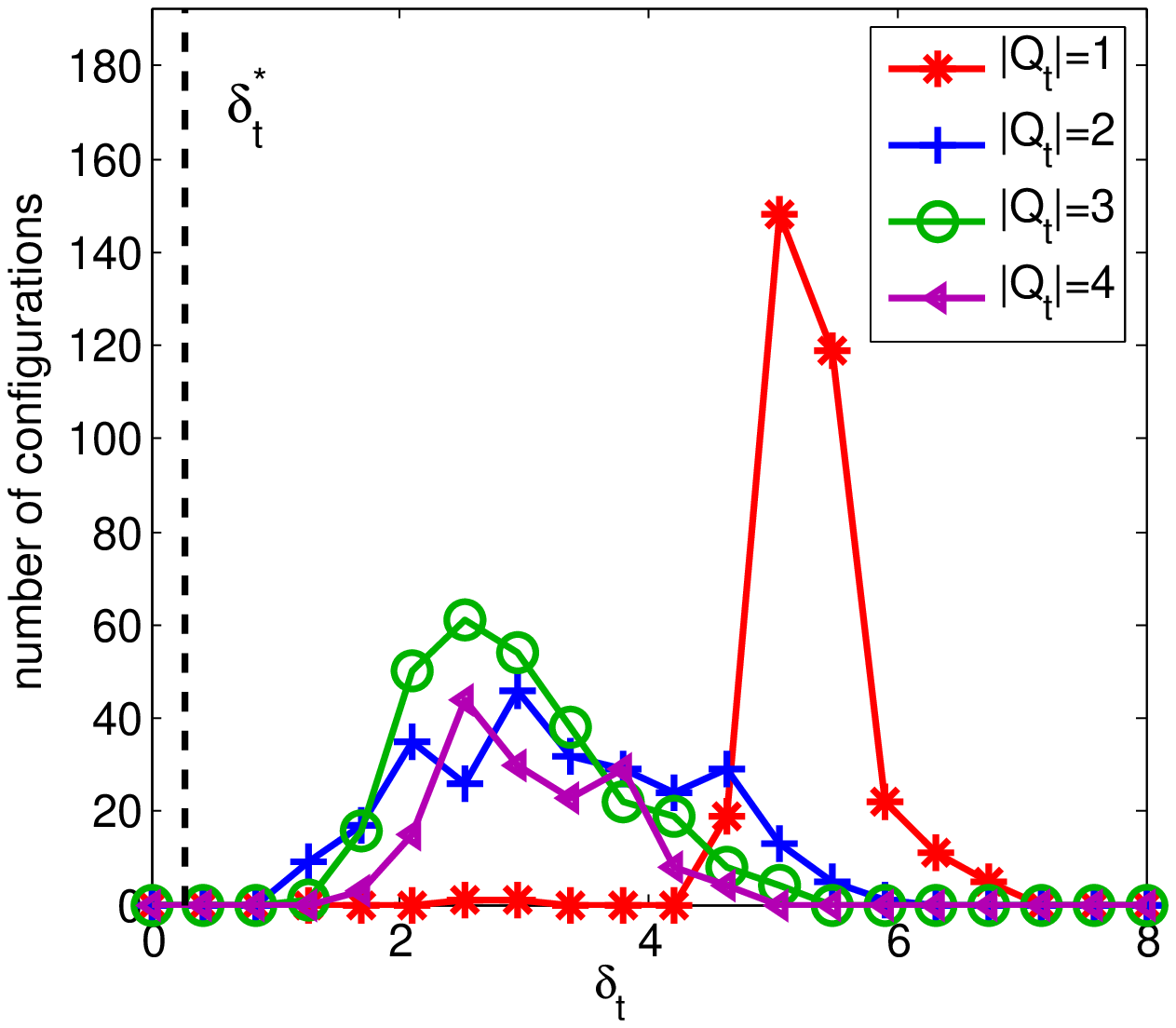}
\includegraphics[height=\fs]{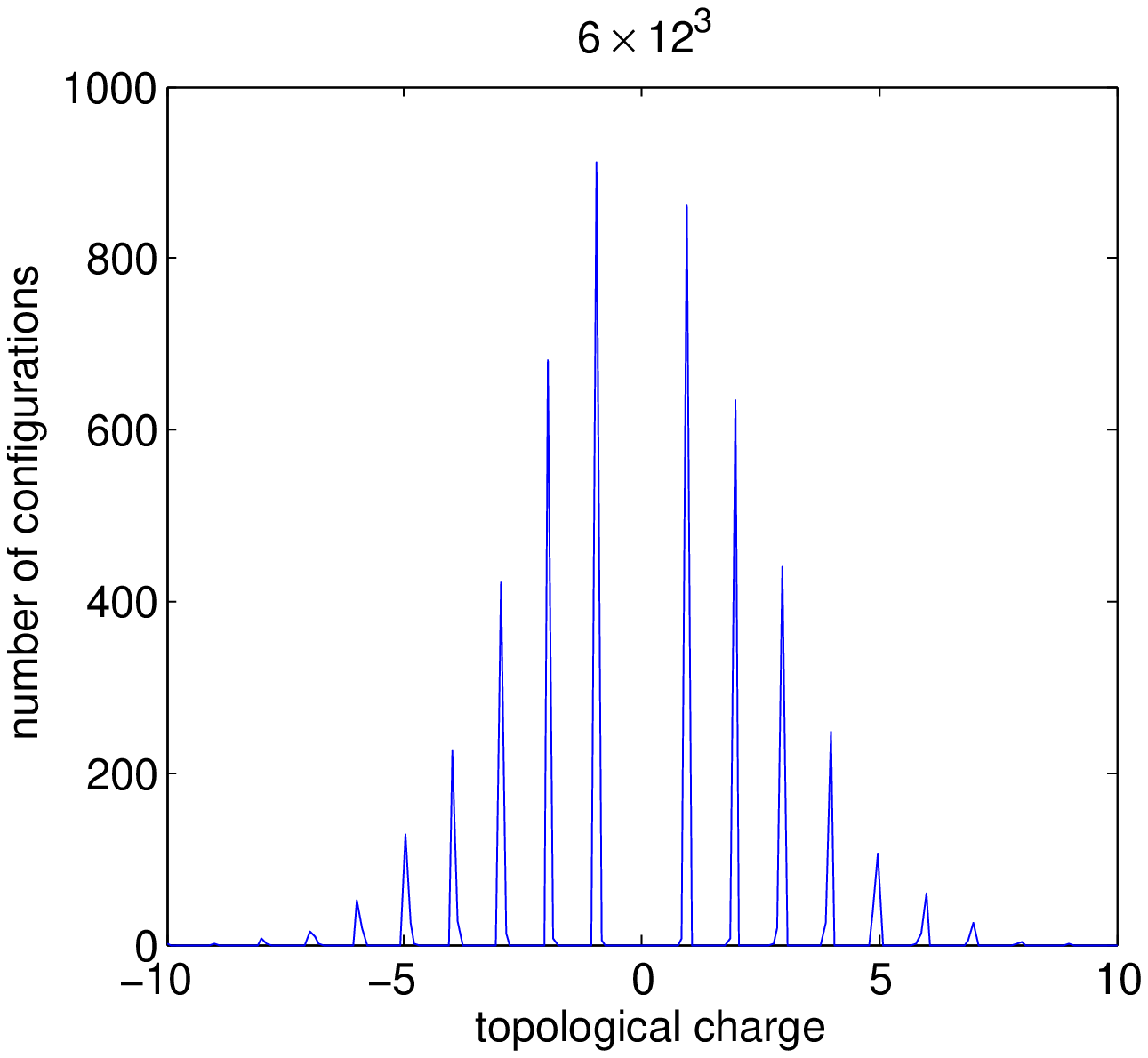}
\includegraphics[height=\fs]{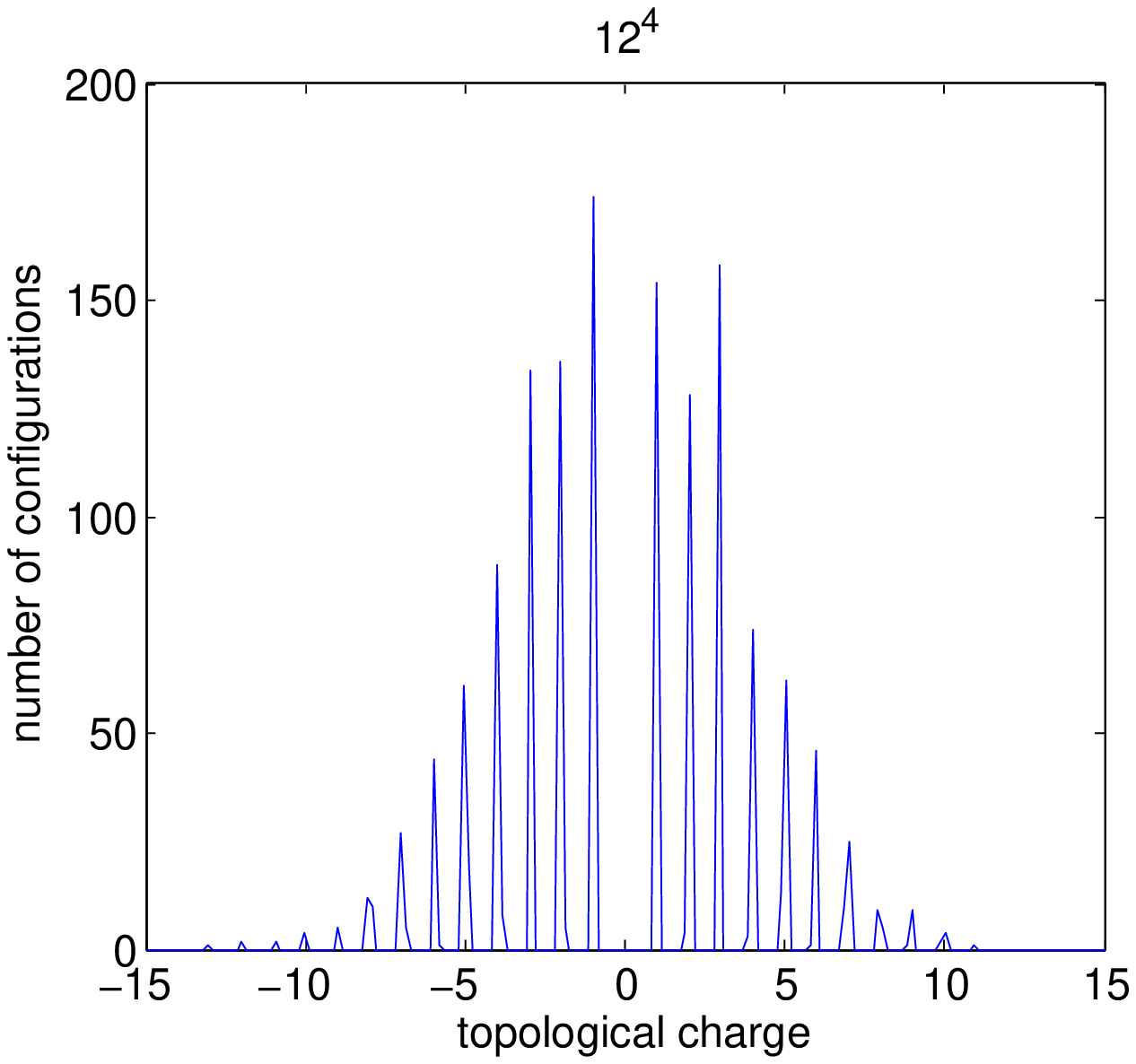}
\caption{Distribution of non-staticity (upper) and topological charges
(lower) for the $6\times 12^3$ (left) and $12^4$ (right) ensembles.}
\label{fig:aspect}
\end{figure}
\noindent
In \fref{fig:avgdist} it becomes clear by comparison that 
the first cluster (a) is well represented by calorons with constituents 
uniformly distributed on the lattice. 
The average distance is comparably large and consequently they are 
static. On the other hand, configurations forming cluster (b) have 
emerged from cooling because of a minimal violation of the equation of 
motion achieved in a later stage of cooling. The constituents are squeezed 
and the action density has become non-static, which can be monitored during 
cooling by increasing values of $\delta_t$.

The correlation between non-staticity and monopole distance 
qualitatively satisfies the analytical expectation while the effect of the
stopping condition reflects the expected, unavoidable side-effect of cooling 
with respect to the standard Wilson action.

To understand what changes when the temperature is lowered at fixed volume
and approaches zero temperature (on the symmetric torus) we have studied
the same properties for an ensemble on a $6\times 12^3$ and a $12^4$
lattice. $N_\tau$ and $N_\sigma$ together (by the aspect ratio of the lattice) 
characterize to what extent finite $T$ in infinite volume, 
$S^1\times \mathbb{R}^3$, is modeled. 
Only if $~N_\sigma \gg N_\tau~$ the caloron finite volume effects 
(caused by overlapping constituents) can be
neglected.  By sending $~N_\tau \to N_\sigma~$ and keeping the size (and 
distance) of the object small compared to the box length, more instanton-like 
classical solution are expected.\\

\begin{figure*}
\centering
\includegraphics[height=1.3\fs]{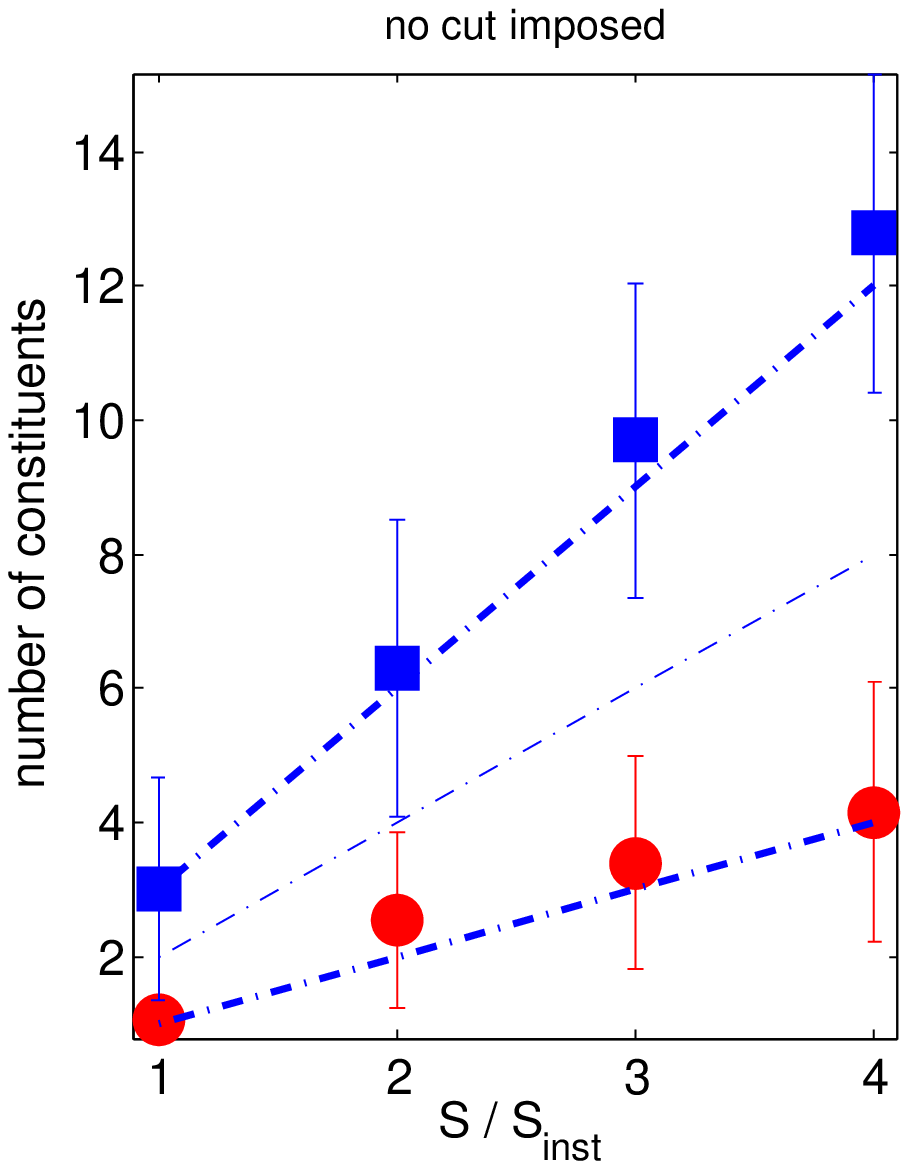}
\includegraphics[height=1.3\fs]{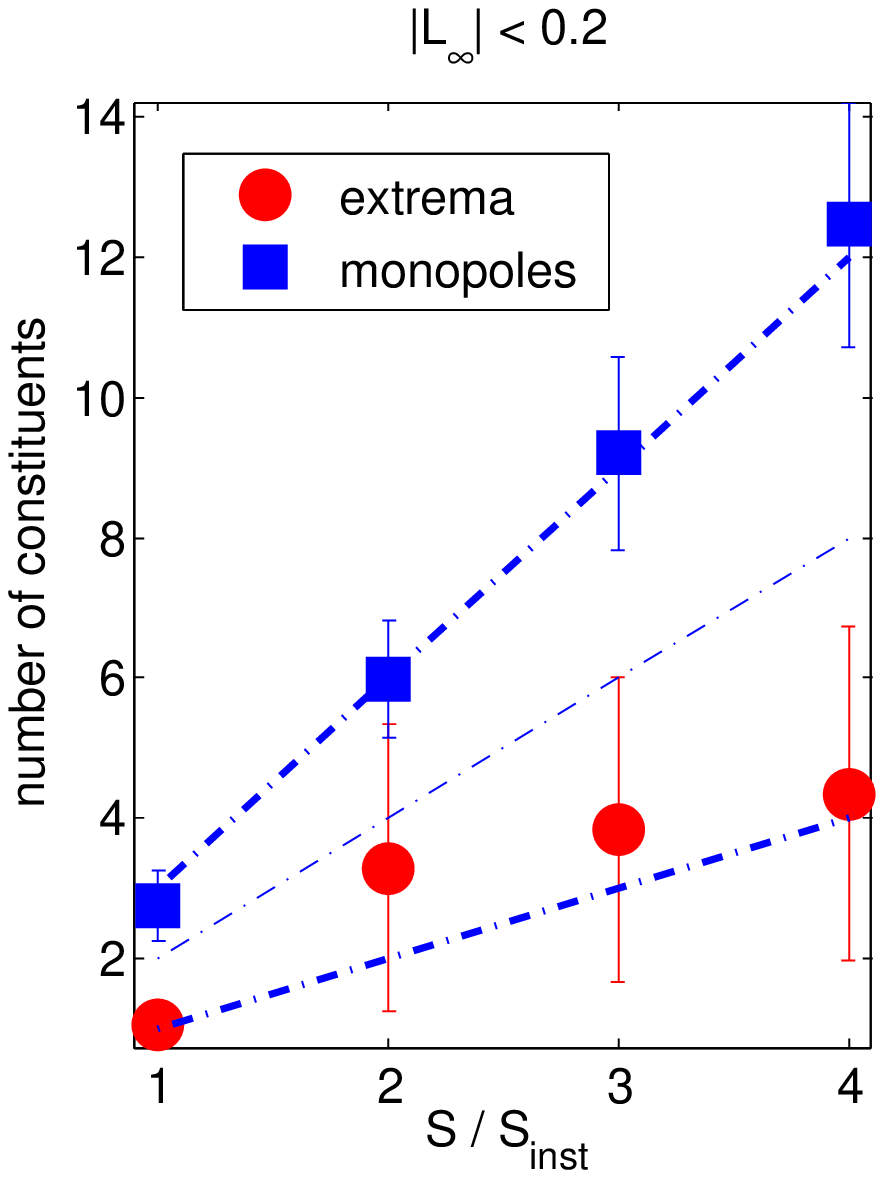}
\includegraphics[height=1.3\fs]{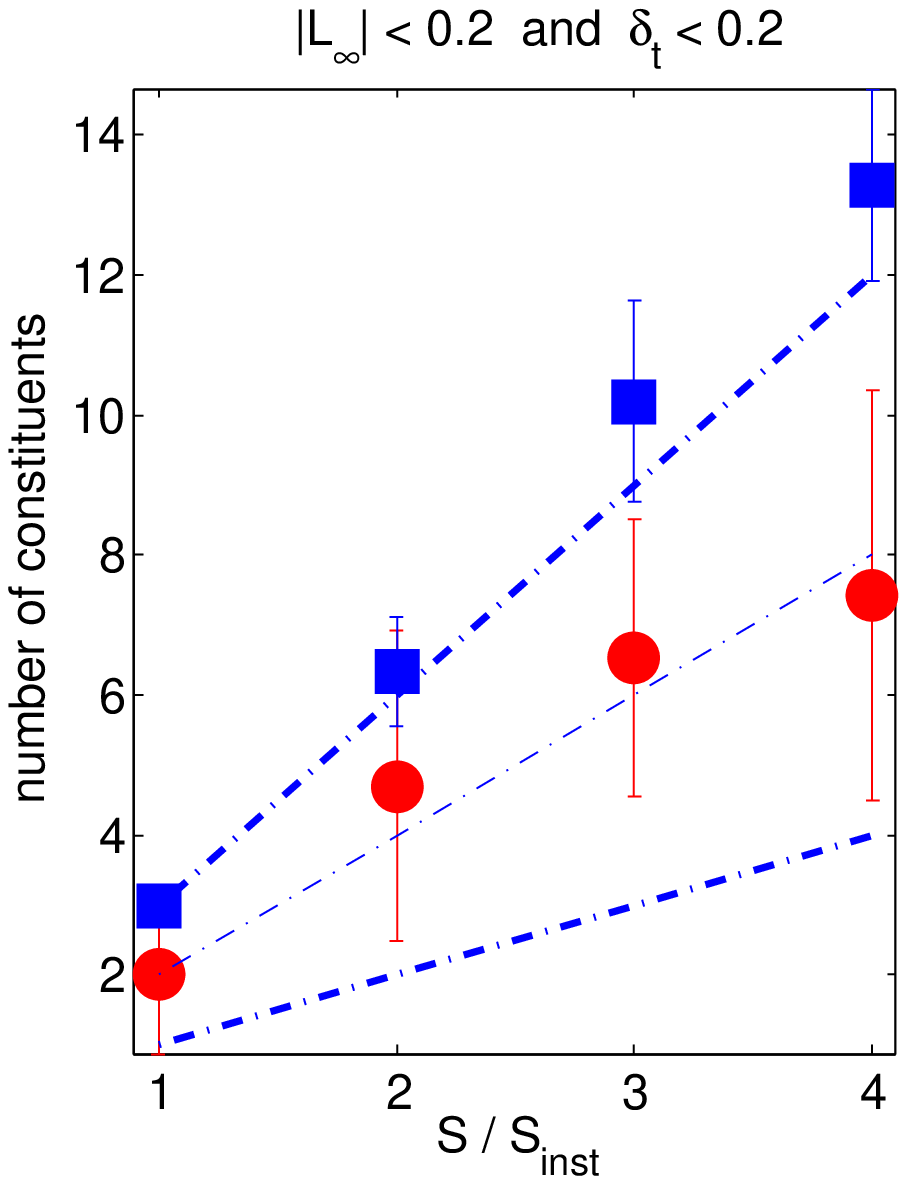}
\caption{Dissociation into constituents from local maxima of action
  and monopole counting (defined by coinciding eigenvalues) for
  $6\times 12^3$ lattice with different cuts imposed.}
\label{fig:ntsixcase}
\end{figure*}

\noindent
In our earlier $SU(2)$ lattice studies a recombination of dissociated
caloron constituents into a single non-dissociated object, 
still with non-trivial holonomy, has been observed  
in the limit $~N_\tau \to N_\sigma~$ \cite{Ilgenfritz:2004ws}~\footnote{A similar 
observation has been made employing the method of ``adiabatic cooling'' 
which uses an anisotropic lattice~\cite{Bruckmann:2004ib}.}.  
We repeat this study for
$SU(3)$ by using lattices with aspect ratios $~N_\tau/N_\sigma=6/12~$ and 
$~N_\tau/N_\sigma=12/12~$ in creating ensembles of configurations 
cooled until the stopping condition \ref{stop2} is satisfied. 
In complete analogy to the $SU(2)$ case static calorons disappear if the 
aspect ratio $~N_\tau/N_\sigma~$ is increased 
(upper part of \fref{fig:aspect}). In \fref{fig:ntsixcase} the number 
of constituents, found for different topological sectors $~|Q_\tx{t}|~$, for
$N_\tau=6$ is shown. Different cuts with respect to the holonomy
$|\Hol|$ and non-staticity $\delta_t$ are applied.
\noindent
Despite the bigger aspect ratio $3~|Q_\tx{t}|$ monopoles are still visible 
whereas a smaller number of local extrema is observed in the
topological charge density, closer to one lump per unit charge, if no
further cuts with respect to holonomy and non-staticity are applied. 
Since the interpretation of the Polyakov loop (corresponding to a
choice of a ``time'' direction) would be ambiguous,
this has not been checked on the symmetric lattice.
We would like to remind the reader that Gattringer and 
Pullirsch~\cite{Gattringer:2004pb} have found on the symmetric torus, 
similar to the observations at $T\ne0$ ~\cite{Gattringer:2002tg},
jumping of the zero-mode for $Q_\tx{t}=\pm 1$ Monte-Carlo configurations
under a change of boundary conditions.
In the light of the present studies we would doubt an interpretation in
terms of jumping between fractionally-charged constituents.

\section{Conclusions}
We have applied cooling to $SU(3)$ Monte Carlo fields in the 
confined phase at finite temperature. 
The typical KvB behavior of these fields was shown for examples with 
$~|Q_\tx{t}|=1,2~$ and $~Q_\tx{t}=0~$ by analyzing field strength, 
Polyakov loop, action and topological charge profiles and the behavior 
of low-lying modes of the Dirac operator. This part was complemented 
by studying the content of the whole ensemble, especially with respect 
to the monopole content of solutions in each topological charge sector. 
The outcome is biased by the choice of lattice action -- we have used 
the standard Wilson action -- which can be compensated by a 
suitable stopping condition for the cooling iteration. 
Still, $~|Q_\tx{t}|=1~$ plays an exceptional role.
However, in agreement with the analytical expectation we find $~3~|Q_\tx{t}|~$
constituent monopoles (for $~|Q_\tx{t}|=1,\dots,4~$) which are defined by the
coincidence of two eigenvalues of $~\Pol(\vec{x})~$.

Zero modes of the Dirac operator with adjustable boundary conditions
are an useful filter to explore the semiclassical structure. 
Chirally improved Dirac operators are necessary 
for the study of its low-lying modes to be applicable 
directly to unsmeared Monte Carlo configurations.
Using the full feature of exploring the spectral flow depending on the boundary 
condition leads to high computational costs such that only small samples 
of configurations can be studied.

Uncovering the topological structure requires one to introduce 
notion of smoothness. Therefore smearing has been applied to explore the 
topological density or the action density. Both observables can give 
valuable information about the lattice configuration 
already after some amount of cooling or smearing. Nevertheless,
one might miss significant features of KvB calorons by relying
only on those. Having the connection between the holonomy and
deconfinement-confinement phase transition in mind, the Polyakov loop
operator is another very important and valuable tool to investigate the
classical properties. The signature of a monopole is that
$~L(\vec{x}) \equiv (1/3)~\tr\Pol(\vec{x})~$ 
is close to the envelope of the range of the Polyakov loop
(see the upper left panel of \fref{fig:q1flow}).
But it is unlikely that this signature can be directly observed 
in an early stage of smearing or cooling.

In future investigations we will use these tools to explore the
properties of finite temperature lattice field theory near the
$SU(3)$ deconfinement--confinement phase transition by applying
different smearing techniques. Lattice studies can hopefully
help to identify the dissociation of calorons with non-trivial holonomy 
if the temperature is lowered below the critical one, in other words,  
to verify the semiclassical results.

\section{Acknowledgements}
We thank P. van Baal, F. Bruckmann, Ch. Gattringer, B. Martemyanov, 
A. Sch\"afer, St. Solbrig and A. Veselov for numerous exciting discussions,
in particular P. van Baal, F. Bruckmann and Ch. Gattringer for a careful 
reading of the manuscript. E.-M. I. and M. M.-P. acknowledge financial support 
by the DFG (FOR 465 / Mu 932/2). D. P. has been supported by the DFG 
Graduate School GRK 271. 

\bibliographystyle{apsrev}


\appendix

\section{Auxiliary fields}

The auxiliary fields for the construction of  $SU(3)$
$~|Q_\tx{t}|=1~$ calorons are \cite{Kraan:1998sn}:
\beas
\psi(x_0,\vec{x})&=&\frac{1}{2}\,\tr(\mc{A}_3\,\mc{A}_2\,\mc{A}_1)
-\cos(2\pi x_0),\\
\mc{A}_m(\vec{x})&=&\frac{1}{r_m}\begin{pmatrix}r_m&|\vec\rho_{m+1}|\\
0&r_{m+1}\end{pmatrix}\begin{pmatrix}{\rm ch_m} & {\rm sh_m}\\ 
{\rm sh_m}&{\rm ch_m}\end{pmatrix}
\eeas
using the notation ${\rm ch_m}=\cosh(2\pi\nu_m r_m)$, 
${\rm sh_m}=\sinh(2\pi\nu_m r_m)$,  
$r_m=|\vec x-\vec y_m|$ and
$\vec\rho_m=\vec y_m-\vec y_{m-1}$. 

The fermionic density is derivable from the Green's function $\hat{f}_x$

\beas
\nn \hat{f}_x(z,z')&=&
\frac{\pi e^{2\pi it(z-z')}}{r_m\psi}
\Big[e^{-2\pi it}\sinh(2\pi r_m(z-z'))\\&+&\langle
v_m(z')|\pi(\mc{A}_{3}\mc{A}_{2}\mc{A}_{1})|w_m(z)\rangle\Big],
\eeas

\noindent with

\beas
v_m^1(z)&=&-w_m^2(z)=\sinh(2\pi r_m(z-\mu_m)), \\
v_m^2(z)&=&\phantom{-}w_m^1(z)=\cosh(2\pi r_m(z-\mu_m)).
\eeas
\vspace{5mm}

The construction of the caloron field $A_{\mu}$ requires the knowledge 
of the following matrix fields $C_\alpha$ and $\phi$:

\beas
(C_\alpha)_{mk}&=&
\zeta_{m}\bar\eta_{\alpha\beta}\;\zeta_{k}^\dagger\,\partial_\beta
\hat{f}_x(\mu_m,\mu_k),\\
\left(\phi^{-1}\right)_{mk}&=&
\delta_{mk}-\zeta_m\zeta_k^\dagger\;\hat{f}_x(\mu_m,\mu_k).
\eeas
\vspace{5mm}

The anti-selfdual 't Hooft symbols are 
$~\bar{\eta}_{\alpha\beta}=\bar{\sigma}_{[\alpha}\sigma_{\beta]}$.
The permutation $~\pi(\mc{A}_{3}\mc{A}_{2}\mc{A}_{1})~$ is
$~\mc{A}_{3}\mc{A}_{2}\mc{A}_{1}$, $~\mc{A}_{1}\mc{A}_{3}\mc{A}_{2}~$
or $~\mc{A}_{2}\mc{A}_{1}\mc{A}_{3}~$ for $~m=1,2$ or $3$, respectively. In the
expression for $~\hat{f}~$ the assumption $~\mu_m\!\leq\!z'\!\leq\!
z\!\leq\!\mu_{m+1}~$ has been made. This can be generalized by 
$~\hat{f}_x(z',z)\!=\!\hat{f}_x^*(z,z')$.
If the constituent positions $~\vec{y}_m~$ lie in the $~x_1-x_2~$ plane, one can use
$~\zeta_m=(|\vec{\rho}_m|,i\rho_m^2-\rho_m^1)/\sqrt{2\pi|\vec{\rho}_m|}$.
In general one has $~\vec{\rho}_m=-\pi\zeta_m\vec{\tau}\zeta_m^\dagger~$
with the Pauli vector $~\vec{\tau}=(\tau_1,\tau_2,\tau_3)$.
The temporal shift $~y_0~$ is set to zero and 
$~b\;=\;1/T\;=\;1~$ throughout these formulas. The angles $~\mu_m~$ 
are determined by the holonomy
$~\Hol\simeq\exp(2\pi i\;\diag(\mu_1,\mu_2,\mu_3))$. 

\end{document}